\documentclass[floatfix,a4paper,aps,prd,twocolumn,preprintnumbers,nofootinbib]{revtex4-1}  
\usepackage{epsfig} 
\usepackage{amsmath} 
\usepackage{amssymb} 
\usepackage{amsfonts} 
\usepackage{graphicx} 
\usepackage{graphics,subfigure} 
\usepackage{float} 
\usepackage{booktabs} 
\usepackage{color} 
\usepackage{geometry} 
\usepackage{rotating} 
\usepackage{siunitx} 
\usepackage{slashed} 
\usepackage{appendix}
\usepackage{hyperref}
\usepackage[utf8]{inputenc}
\usepackage[normalem]{ulem}
 
\newcommand{\lsim}{\raisebox{-0.13cm}{~\shortstack{$<$ \\[-0.07cm] $\sim$}}~} 
\newcommand{\gsim}{\raisebox{-0.13cm}{~\shortstack{$>$ \\[-0.07cm] $\sim$}}~} 
 
\newcommand{\bea}{\begin{eqnarray}} 
\newcommand{\eea}{\end{eqnarray}} 
\newcommand{\beq}{\begin{equation}} 
\newcommand{\eeq}{\end{equation}} 
\newcommand{\beqa}{\begin{eqnarray}} 
\newcommand{\eeqa}{\end{eqnarray}} 
\newcommand{\bit}{\begin{itemize}} 
\newcommand{\eit}{\end{itemize}}

\newcommand{\gev}[1]{\SI{#1}{\giga\electronvolt}} 
\newcommand{\mev}[1]{\SI{#1}{\mega\electronvolt}} 
\newcommand{\tev}[1]{\SI{#1}{\tera\electronvolt}}

\newcommand{\ifb}[1]{\SI{#1}{\femto\barn^{-1}}} 
 
\newbox\charbox 
\newbox\slabox 
\def\s#1{{      
    \setbox\charbox=\hbox{$#1$} 
    \setbox\slabox=\hbox{$/$} 
    \dimen\charbox=\ht\slabox 
    \advance\dimen\charbox by -\dp\slabox 
    \advance\dimen\charbox by -\ht\charbox 
    \advance\dimen\charbox by \dp\charbox 
    \divide\dimen\charbox by 2 
    \raise-\dimen\charbox\hbox to \wd\charbox{\hss/\hss} 
    \llap{$#1$} 
}}

\begin{document}

\title{Prospects for natural SUSY} 
 
\author{J.~S.~Kim} 
\email[]{jong.kim@csic.es} 
\affiliation{Instituto de F\'{\i}sica Te\'{o}rica UAM/CSIC, Madrid, Spain}  

\author{K.~Rolbiecki} 
\email[]{krolb@fuw.edu.pl} 
\affiliation{Institute of Theoretical Physics, University of Warsaw, Poland}  
 
\author{R.~Ruiz} 
\email[]{rruiz@ific.uv.es} 
\affiliation{Instituto de F\'{\i}sica Corpuscular, IFIC-UV/CSIC, Valencia, Spain}

\author{J.~Tattersall} 
\email[]{tattersall@physik.rwth-aachen.de} 
\affiliation{Institute for Theoretical Particle Physics and Cosmology, RWTH Aachen, Germany}
 
\author{T.~Weber} 
\email[]{torsten.weber@rwth-aachen.de} 
\affiliation{Institute for Theoretical Particle Physics and Cosmology, RWTH Aachen, Germany} 
 
\begin{abstract}

As we anticipate the first results of the 2016 run, we assess the discovery potential of the LHC to 
``natural supersymmetry''. To begin with, we explore the region of the model parameter space that can be excluded
with various centre-of-mass energies (13~TeV and 14~TeV) and different luminosities (20~fb$^{-1}$, 100~fb$^{-1}$,
300~fb$^{-1}$ and 3000~fb$^{-1}$). We find that the bounds at 95\% C.L. on stops vary 
from $m_{\tilde{t}_1}\gsim800$~GeV expected this summer to $m_{\tilde{t}_1}\gsim1500$~GeV at the end 
of the high luminosity run, while gluino bounds are expected to range from $m_{\tilde{g}}\gsim1700$~GeV
to $m_{\tilde{g}}\gsim2500$~GeV over the same time period. However, more pessimistically, we find that if
no signal begins to appear this summer, only a very small region of parameter space
can be discovered with 5$\sigma$ significance. For this conclusion to change, we find that both
theoretical and systematic uncertainties will need to be significantly reduced.

\end{abstract} 
 
\preprint{IFT-16-054}
\preprint{TTK-16-17 } 

\maketitle 
  
\section{Introduction} 
\label{sec:intro} 

The Large Hadron Collider (LHC) has entered the next stage of its endeavour to 
discover physics beyond the standard model (BSM). Unfortunately, the lack of a clear signal 
of new physics has already significantly constrained the allowed parameter space of 
many BSM  models. 
In particular, the limits on supersymmetric (SUSY) particles have moved, 
in many cases, well beyond 1~TeV (see e.g.~\cite{Aad:2015baa,Khachatryan:2016nvf,Aad:2015pfx,Aad:2015iea,Chatrchyan:2013iqa,Khachatryan:2016oia}).    

The improved limits on SUSY seem to be particularly damaging for constrained 
models that only have a few free parameters and do not offer much 
flexibility in the resulting mass spectra \cite{Chamseddine:1982jx}. In addition, the heavy 
particles deduced from the limits threaten one of the main motivations 
for SUSY, which is stabilisation of the electroweak symmetry breaking 
scale (EWSB). However, the constrained models rely on assumptions 
about the details of the unknown SUSY breaking mechanism. Once we relax the 
assumptions about the high energy completion of the theory, the parameter 
space opens up to many new possibilities that could accommodate 
relatively light, but still not excluded, particles, while 
addressing some of the problems related to the naturalness of EWSB. 

It has been noted~\cite{Feng:1999mn,Kitano:2006gv,Baer:2012uy,Baer:2011ec,Papucci:2011wy} that for a successful 
stabilisation of the 
electroweak scale, only a small subset of the many SUSY states are required to be light, and
consequently, this helps to relax the general bounds on SUSY models with relatively light spectra. 
Such observations have
renewed interest in so-called ``natural supersymmetry'' models in the context of LHC 
searches. The motivation behind natural supersymmetry is the question of fine-tuning, 
see e.g.~\cite{Barbieri198863,Ellis:1986yg}, which is required to obtain the desired 
scale of EWSB. Low fine-tuning demands certain parameters to be of order of the 
EWSB scale, in particular, the supersymmetric partners of the top quarks and Higgs bosons are 
expected to be light. A detailed analysis 
also reveals~\cite{Casas:2014eca} that the amount of fine-tuning is 
closely related to the mass of the gluino, the superpartner of the gluon. Thus, within 
the naturalness paradigm, one can expect that SUSY is within the LHC reach, and many
studies have investigated the phenomenology of natural SUSY at the LHC 
(see e.g.~\cite{Han:2013kga,Kowalska:2013ica,Buchmueller:2013exa,Baer:2012uy,Belanger:2015vwa,Kobakhidze:2015scd,Kim:2015dpa,Drees:2015aeo}). On 
the other hand, the measured mass of the Higgs boson~\cite{Aad:2015zhl} requires 
rather heavy partners of top quarks~\cite{Bechtle:2012jw} in the minimal supersymmetric standard model (MSSM) which clearly already puts the naturalness
paradigm into question.  

In Ref.~\cite{Drees:2015aeo} the authors presented constraints on a natural supersymmetry scenario derived from Run 1 results at the LHC. 
They called their scenario the minimal ``natural supersymmetry'', since only the superparticles which were
required to cancel the leading quadratically divergent corrections to the standard model (SM) Higgs mass were
kept light. They derived robust limits, implying that the lighter stop mass eigenstate with mass below 230~GeV or a gluino with mass below 440~GeV is always excluded.

A dedicated analysis targeting natural SUSY was carried out by ATLAS in Ref.~\cite{Aad:2015pfx}. It
resulted in rather weak limits, both in the simplified model setup and in a natural-supersymmetry-motivated phenomenological MSSM in a three-dimensional parameter space.
The excluded stop masses were typically in the range $350$--$500$~GeV, when the 
higgsino mass $\lsim \gev{150}$. If simultaneous production
of stops and sbottoms was taken into account, stop masses up to $\gev{680}$ were excluded 
for higgsinos at the current limit ${\sim}\gev{100}$.

In this paper we revisit prospects for discovery of natural SUSY in the 
Run 2 of the LHC as well as after the future high luminosity (HL) upgrade. We 
follow a staged approach, which means that we consider different integrated 
luminosities and centre-of-mass energies that are currently planned at CERN~\cite{lhc-prospects,lhc-prospects2}:
\begin{itemize}
    	\item 13~TeV, 20~fb$^{-1}$ -- expected for 2016;
	\item 13~TeV, 100~fb$^{-1}$ -- Run 2 total luminosity by 2018;
	\item 14~TeV, 300~fb$^{-1}$ -- Run 3 luminosity, 2020--2023;
	\item 14~TeV, 3000~fb$^{-1}$ -- HL upgrade. 
\end{itemize}
For each stage we analyse the expected exclusion limits, and in the case of the high luminosity
upgrade we also consider the discovery reach for natural SUSY. 

One of the main motivations of the current study is the question of what the potential future 
discovery reach is with increased centre-of-mass energy and integrated luminosity assuming 
that no signal is observed this year after collecting $20$--$30$~fb$^{-1}$ of data. This 
is a relevant issue since 95\% confidence level (C.L.) exclusion limits advance much 
faster than the discovery reach that conservatively requires 5$\sigma$ significance 
for serious consideration. Therefore, if null results are obtained this year, the question is, can
we still hope to make a discovery at the LHC? We find that the answer has a significant dependence 
on how well systematic uncertainties can be reduced in the future and we therefore 
consider different scenarios for their future evolution.
 
In this study we closely follow the numerical procedure of Ref.~\cite{Drees:2015aeo}. We investigate the minimal six-dimensional natural SUSY model 
where higgsinos, gluinos and third-generation squarks are all assumed to 
be in the vicinity of the TeV scale. We work within the MSSM setup, with the lightest 
neutralino as the lightest supersymmetric particle (LSP) assuming 
conserved $R$-parity~\cite{Ibanez:1991pr}. We have randomly generated 20000 benchmark points in the minimal natural supersymmetry parameter space. 
The numerical analysis is performed with  
simulated Monte Carlo (MC) samples further processed by \texttt{CheckMATE}~\cite{Drees:2013wra}
using \texttt{Delphes}~\cite{deFavereau:2013fsa}  
for detector simulation and analysis. We include several ATLAS searches that 
are suited particularly well for natural SUSY, both inclusive and more specific 
for this model. The searches were optimised for different stages of the LHC operation 
from low to high luminosities and an increased centre-of-mass energy.  

The remainder of this article is organised as follows. In Sec.~\ref{sec:natural_susy_setup} we 
describe our natural SUSY inspired scenario and its collider signatures. Next, in 
Sec.~\ref{sec:numerical_analysis}, we first discuss the numerical tools employed for this 
study and then discuss the setup for our scan. Finally, we present our numerical results in Sec.~\ref{sec:results} and show limits on the stop and the gluino mass as a function of the LSP mass. We conclude in Sec.~\ref{sec:summary}.

\section{Natural Supersymmetry Setup} 
\label{sec:natural_susy_setup}
The electroweak symmetry breaking condition of the MSSM dictates the following relation,
\beq\label{eq:naturalness}
\mu^2=\frac{m_{H_d}^2-m_{H_u}^2\tan^2\beta}{\tan^2\beta-1}-\frac{1}{2} M_Z^2,
\eeq
which correlates the supersymmetric Higgs mixing parameter $\mu$ with the soft supersymmetry 
breaking mass terms $m_{H_d}^2$ and $m_{H_u}^2$, the ratio of the two Higgs vacuum expectation 
values $\tan\beta=\frac{v_u}{v_d}$ and the observed $Z$ mass. This tree level equation serves 
as the starting point for a qualitative discussion of naturalness. We regard the MSSM as natural 
if the individual terms are of the same order as $M_Z$. If the supersymmetric particles are
too heavy, they will give large contributions 
to the various mass terms of Eq.~(\ref{eq:naturalness}).
As a consequence, all individual contributions have to be finely tuned in order to obtain the
correct $Z$ boson mass. 

However, not all supersymmetric sparticles contribute equally to Eq.~(\ref{eq:naturalness}). The most 
important contribution comes from the left-hand side, i.e., the $\mu$ parameter. Hence, a very 
simple definition of naturalness only requires the $\mu$ parameter to be of the order of the
electroweak scale, and consequently, this implies 
light higgsinos~\cite{Baer:2012uy,Baer:2011ec}. For example, the MSSM with $|\mu|\le M_Z$ would 
be regarded as {\it very} natural, although this is already in conflict with the LEP2 
results~\cite{LEPconst}. Recently, it has been pointed
out that an exception to this rule is possible if an additional Higgsino soft breaking term
is included \cite{Ross:2016pml}, and this allows for $m_{\tilde{H}}\lesssim 500$~GeV for ${\sim}10$\% tuning. 

Unfortunately, radiative corrections complicate this picture
significantly and must be included for consistency since they can lead to very large effects. 
The dominant one-loop corrections to $m_{H_u}^2$ and $m_{H_d}^2$ in Eq.~(\ref{eq:naturalness}) 
are driven by the top partners (stops) since they couple with the large top Yukawa coupling 
strength. As the one-loop 
contribution scales with the square of the stop masses, the stops should not be too heavy 
($m_{\tilde{t}}\lesssim 800$~GeV for ${\sim}10$\% tuning \cite{Papucci:2011wy})
in order to prevent unduly large contributions to the right-hand side of Eq.~(\ref{eq:naturalness}). 
Analogously, two-loop corrections renormalise the Higgs soft breaking masses as well and put 
an upper limit on the gluino mass ($m_{\tilde{g}}\lesssim 1200$~GeV for ${\sim}10$\% 
tuning~\cite{Papucci:2011wy}), though the constraints are weaker than those present 
on the higgsino and stop mass. The following qualitative picture emerges from the naturalness 
principle: light higgsinos, not too heavy stops and gluinos that are not 
decoupled from the LHC phenomenology~\cite{Feng:1999mn,Papucci:2011wy,Kitano:2006gv}. 

However, the exact definition of naturalness depends on the degree of fine-tuning. A 
quantitative measure of fine-tuning was defined in \cite{Barbieri198863}:
\begin{equation}\label{eq:finetuning}
\Delta=\left|\frac{a}{M_Z^2}\frac{\partial M_Z^2}{\partial a}\right|,
\end{equation}
where $a$ is an input parameter which can be a soft breaking parameter at the 
electroweak scale, but it could equally well be a universal soft breaking parameter 
at some high scale of a constrained supersymmetric model. A value of $\Delta=20$ would 
then correspond to a fine-tuning of $\Delta^{-1}=5\%$. Equation~(\ref{eq:finetuning}) together 
with the amount of fine-tuning $\Delta$ that one is willing to accept, gives an upper limit on 
the sparticle masses. Due to their much smaller couplings, it is clear that the first two generation squarks and sleptons, as well as the electroweak bino and wino, only have a negligible impact on naturalness.

\begin{figure} 
\includegraphics[clip, trim=0cm 5cm 0.cm 0cm,width=0.5\textwidth]{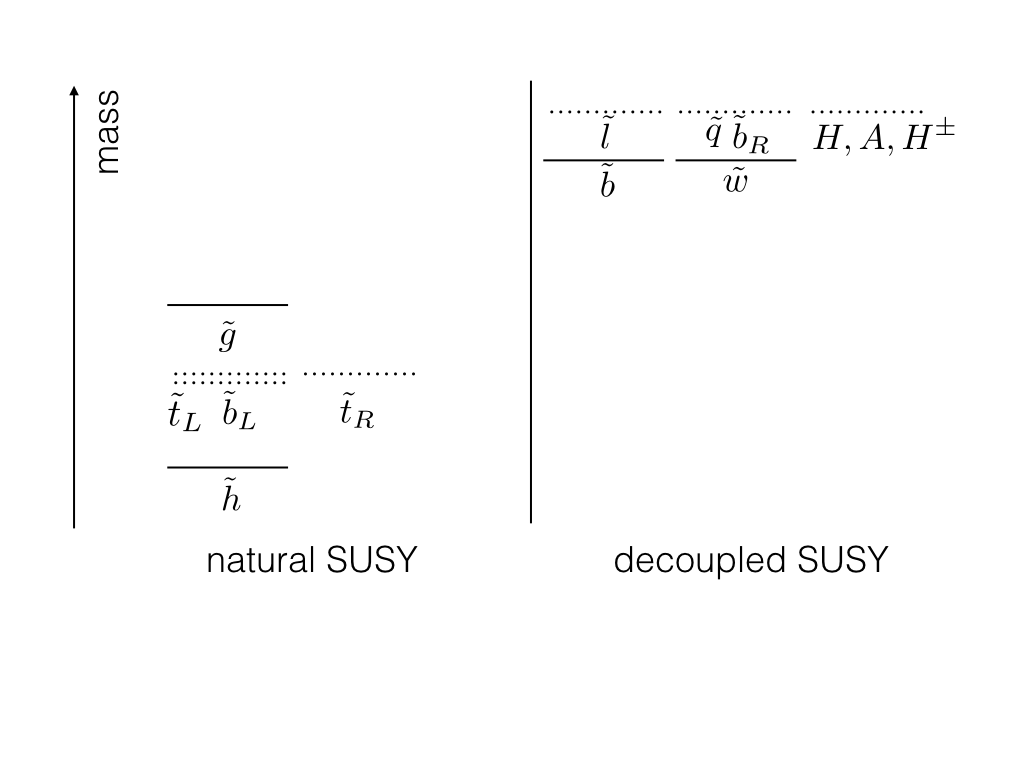} 
\caption{The minimal natural SUSY mass spectrum are shown on the left, while the remaining supersymmetric particles are decoupled on the right.}
\label{fig:nsusy_mass} 
\end{figure}

Here, we want to be agnostic about the exact amount of {\it natural} fine-tuning, since this is somewhat subjective. We 
closely follow the description of the natural SUSY particle spectrum discussed in Ref.~\cite{Drees:2015aeo}. We only consider supersymmetric particles which are essential to cancel the quadratic divergences to the Higgs mass, and thus our scenario can be regarded as the minimal natural MSSM. We demand higgsinos with masses below the TeV scale, third-generation soft squark masses up to 2 TeV and gluinos with masses below 3 TeV. We assume that all sleptons and the first- and second-generation squarks have a common mass scale at $m_{\tilde f}^2=1.5\times10^7$ GeV$^2$. Since we only consider scenarios with $\tan\beta\le20$, the sbottom loop contribution to the Higgs mass is negligible, and thus we fix the mass of the right-handed sbottom $b_R$ at the same common sfermion mass scale $m_{\tilde f}$. However, since we consider both stop states, in particular, the SU(2)-doublet stop, we also have to include the left-handed sbottom due to weak isospin invariance. $A_b$ does not play a role in the phenomenology, and we fix it to $A_b=0$ GeV without loss of generality. Finally, we set the bino mass parameter $M_1$ and wino mass parameter $M_2$ to 3 TeV. All additional MSSM Higgs bosons are assumed to be decoupled by fixing the pseudoscalar mass at $m_A=2.5$ TeV. Figure~\ref{fig:nsusy_mass} shows the supersymmetric particle spectrum of the minimal natural MSSM. In summary, our natural SUSY scenario is parametrised in terms of six parameters: the supersymmetric 
Higgs mixing parameter $\mu$, the gluino mass parameter $M_3$, $\tan\beta$, the third-generation SU(2)-doublet squark soft-breaking parameter $m_{Q_t}$, the corresponding SU(2)-singlet soft-breaking parameter $m_{t_R}$, and the top trilinear soft-breaking term $A_{t}$.

In natural SUSY the dominant production mechanism is the pair production of third-generation sparticles and gluinos via strong interactions,
\beq\label{eq:production}
pp\rightarrow \tilde g \tilde g,\quad pp\rightarrow \tilde t_{1(2)}\tilde t_{1(2)}^*,\quad pp\rightarrow \tilde b_1 \tilde b_1^*.
\eeq
Here, we will not consider direct higgsino pair production. In general, the mass splitting between the higgsino mass eigenstates is a few GeV, and thus the energy release of the SM particles in $\tilde\chi_1^\pm$ and $\tilde\chi_2^0$ decays is too small to be detected at the LHC. Both are invisible at the LHC provided their decays into the $\tilde \chi_1^0$ LSP are prompt. As a result, only higgsino pair production in association with a jet yields a viable collider signature. However, the small production cross section for the monojet final state, together with the large systematic uncertainty of the SM background prediction,\footnote{For a discussion of the systematic uncertainty, please refer to the discussion in Ref.~\cite{Kim:2015hda}} allows only the exclusion of very light higgsinos even at the high luminosity phase of the LHC \cite{PhysRevD.89.055007}. 

The decay modes of the coloured sparticles are complicated and depend on the mass hierarchy as well as the mixing of the third-generation sparticles. If the gluino is decoupled, the following third-generation scalar decays can occur,
\bea
\tilde t_a \rightarrow t \tilde\chi_l^0,\quad b\tilde\chi_1^\pm,\qquad l=1,2,\label{eq:stopdec1}\\
\tilde b_1 \rightarrow b \tilde\chi_l^0,\quad t\tilde\chi_1^\pm,\qquad l=1,2.\label{eq:stopdec2}
\eea
When we neglect phase-space suppression, the branching ratios of the stops are determined
by the dominant coupling character of the state in question. Namely, the
SU(2)-doublet decays mainly to $t\tilde \chi_{1,2}^0$ since 
the $\tilde t_L t \tilde \chi_{1,2}^0$ coupling is proportional to the top Yukawa coupling, 
while $\tilde t_L b \tilde \chi_1^\pm \sim Y_b$ in the higgsino case. On the other hand, the 
SU(2)-singlet $\tilde t_1$ decays to $b \tilde \chi_1^+$ and $t\tilde \chi_{1,2}^0$ at similar 
rates, with all couplings proportional to $Y_t$ for higgsinos (see e.g.~\cite{Rolbiecki:2009hk}).

\begin{figure} 
\includegraphics[clip, trim=0cm 5cm 0.cm 0cm,width=0.5\textwidth]{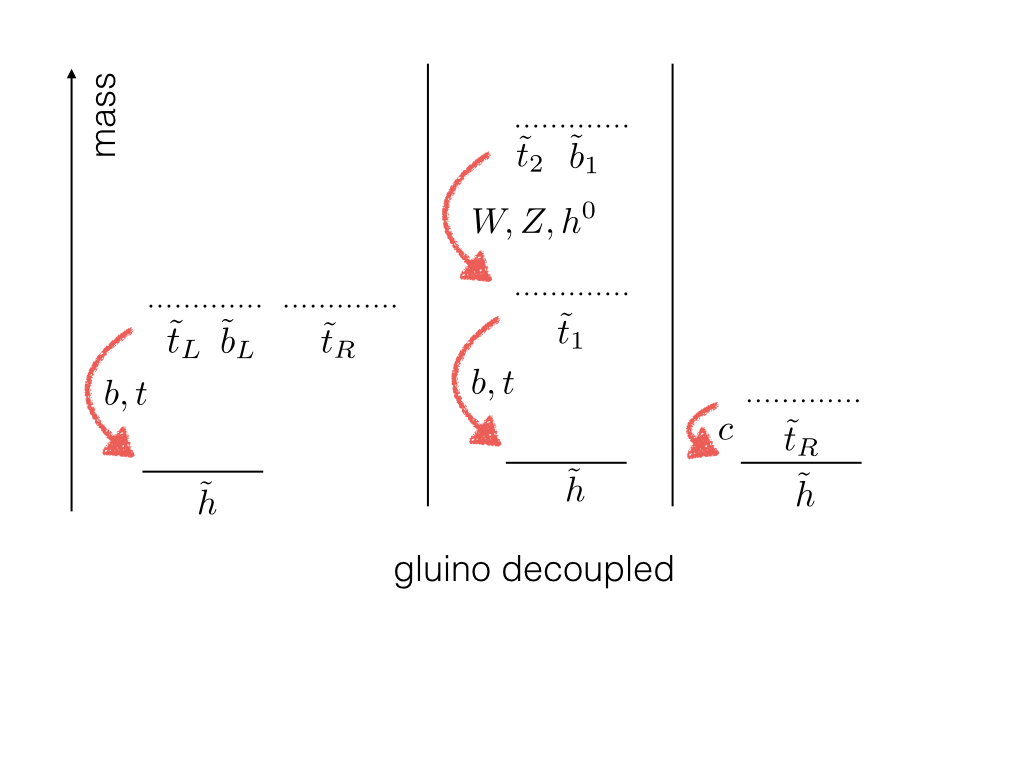} 
\caption{Decay modes in natural SUSY assuming decoupled gluinos.}
\label{fig:decay_modes_nsusy1} 
\end{figure}

Decays into lighter third-generation scalars involving SM gauge bosons and the Higgs scalar are also possible,
\bea
&\tilde t_a \rightarrow \tilde b_1 W^\pm, \quad &\tilde b_1 \rightarrow \tilde t_1 W^\pm,\label{eq:heavy_stop_decay1}\\
&\tilde t_2 \rightarrow \tilde t_1 Z,\ \ \ \quad &\tilde t_2 \rightarrow \tilde t_1 h.
\label{eq:heavy_stop_decay2}
\eea
If the two-body decay modes are kinematically closed, in particular, $\tilde t_1\rightarrow b \tilde \chi_1^\pm$, the stop can have a 
sizeable branching ratio in three- and four-body final states. If all tree-level decays of the lighter stop $\tilde t_1$ are heavily suppressed, the loop-induced decay $\tilde t_1\rightarrow c \tilde \chi_1^0$ can 
be the dominant decay mode of the lighter stop \cite{Hikasa:1987db}. Examples of a typical spectrum in the gluino decoupled
case with the corresponding decay modes are shown in Fig.~\ref{fig:decay_modes_nsusy1}. 

If the gluino is light enough, direct decays of the third-generation sparticles via strong interactions are possible,
\beq
\tilde t_a\rightarrow t\tilde g,\quad \tilde b_1\rightarrow b \tilde g,\qquad a=1,2.
\eeq
However, in this region of parameter space the far larger gluino production rate means that
the production of the heavier stops is phenomenologically unimportant.

\begin{figure} 
\includegraphics[clip, trim=0cm 5cm 0.cm 0cm,width=0.5\textwidth]{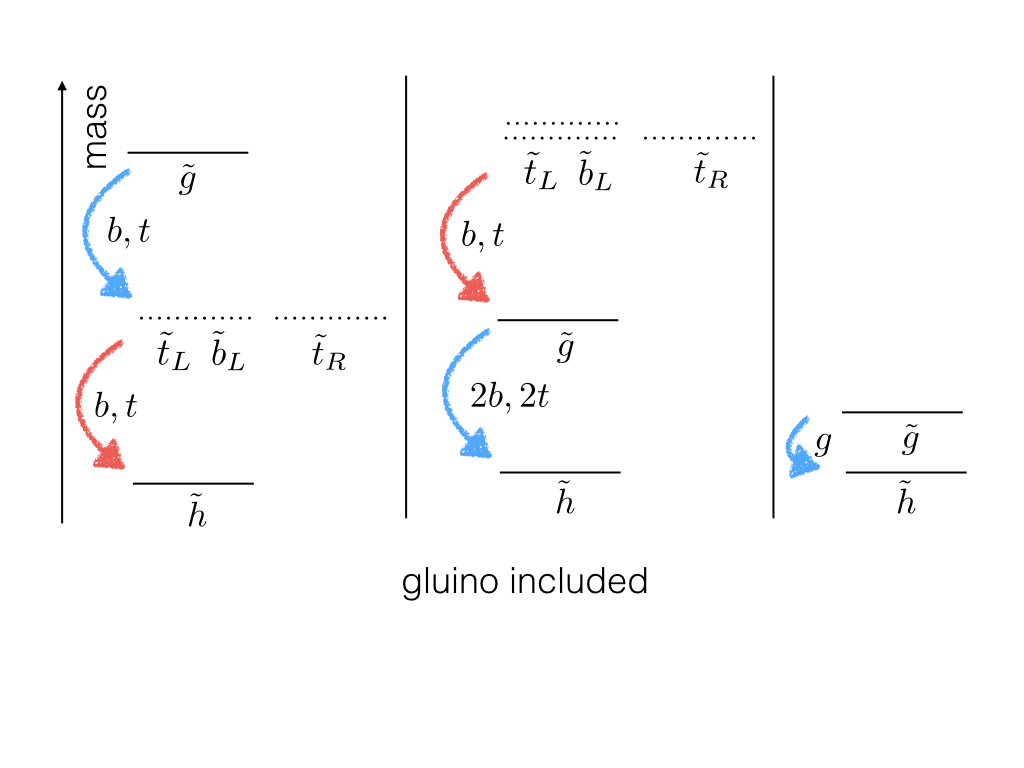} 
\caption{Decay modes in natural SUSY assuming light gluinos in the spectrum.}
\label{fig:decay_modes_nsusy2} 
\end{figure}

Finally, gluinos typically decay via
\beq
\tilde g \rightarrow \tilde t_a \bar t,\quad \tilde t_a^* t,\quad \tilde b_1 \bar b,\quad\tilde b_1^* b,\qquad a=1,2.
\eeq
If the two-body decays are phase-space suppressed, decays via off-shell squarks are possible,
\beq
\tilde g\rightarrow t \bar t\tilde \chi_{1(2)}^0,\quad b \bar b\tilde \chi_{1(2)}^0,\quad \bar t b\tilde\chi_1^+,\quad\bar b t\tilde\chi_1^-.
\eeq
We can already see that the gluino pair production will yield high $b$-jet multiplicities in the final state.
If all third-generation sparticles are decoupled, the loop-induced decay can become dominant \cite{Baer:1990sc},
\beq
\tilde g \rightarrow g\tilde\chi_1^0.
\eeq
Several characteristic decay modes with sample mass hierarchies are summarised in Fig.~\ref{fig:decay_modes_nsusy2}. 

\section{Numerical analysis\label{sec:numerical_analysis} } 

\subsection{Numerical tools} 
\label{subsec:numerical_tools}
The masses, couplings and branching ratios of all sparticles in our natural SUSY setup 
are calculated with {\tt SPheno 3.2.4} \cite{Porod:2011nf}.\footnote{For the Higgs mass constraint we
use {\tt SPheno 3.3.8}.} For each model point, MC 
events are generated with {\tt Pythia 8.210}~\cite{Sjostrand:2014zea,Desai:2011su} at the centre-of-mass 
energy \tev{14} using its default parton distribution function set \cite{Nadolsky:2008zw}. The production 
cross sections are normalised with 
{\tt NLLFAST 4.1}~\cite{Beenakker:1996ch,Beenakker:1997ut,Kulesza:2008jb,Kulesza:2009kq,Beenakker:2010nq,Beenakker:2011fu},  
which computes hadronic SUSY cross sections at next-to-leading order and resums the soft gluon emission at next-to-leading 
logarithmic accuracy. The hadronic events have been passed to {\tt CheckMATE 1.2.1}~\cite{Drees:2013wra,Kim:2015wza,webpage}, 
which is based on the fast detector simulation {\tt Delphes 3.10} \cite{deFavereau:2013fsa} and {\tt Fastjet 3.0.6} for the 
jet reconstruction \cite{Cacciari:2005hq,Cacciari:2008gp,Cacciari:2011ma}. {\tt CheckMATE} tests all model points against experimental searches at the LHC. 
For each model point, it determines the number of expected signal events in all signal regions. 
In order to test the future exclusion sensitivity and discovery potential,  
we compare the signal predictions with the SM background estimates and errors 
provided by the ATLAS Collaboration for a given luminosity and centre-of-mass energy. In 
order to make predictions at other energies or total integrated luminosities, the 
background numbers and errors have to be rescaled accordingly. The precise procedure 
we use is described in Sec.~\ref{subsec:rescaling}.
 
\subsection{Scan procedure}
We have randomly generated numerical values for the natural SUSY parameters listed together with their scan ranges:
\bea
 0.1\,{\rm TeV}\leq&|\mu|&\leq 1.0\,{\rm TeV},\nonumber\\
 0.1\,{\rm TeV}\leq& m_{\tilde Q_t}&\leq 2.0\,{\rm TeV},\nonumber\\
 0.1\,{\rm TeV}\leq& m_{\tilde t_R}&\leq 2.0\,{\rm TeV},\nonumber\\
 0.1\,{\rm TeV}\leq&|M_3|&\leq 3.0\,{\rm TeV},\nonumber\\
&|A_t|&\leq 3.0\,{\rm TeV},\nonumber\\
 1\leq&\tan\beta&\leq 20.\nonumber
\eea
All probability distribution functions for the input parameters are flat, and their lower limits are motivated by the null results from searches for supersymmetric particles performed by the LEP2 and Tevatron experiments. 

The soft breaking parameters are given in the $\overline{\rm{DR}}$ scheme and passed to the spectrum 
generator {\tt SPheno}, which calculates on-shell masses and decay widths as well as low-energy electroweak precision observables. We apply a large number of theoretical and experimental constraints on our benchmark points, which we will discuss in the following. 

We demand that all benchmark points satisfy the conditions for correct electroweak symmetry 
breaking and that the spectrum is tachyon free. Moreover, we demand a CP-even Higgs 
boson with a mass $m_h=125\pm3$ GeV with SM-like couplings. We also test our benchmark points 
against electroweak precision observables such as the $\rho$ parameter~\cite{Drees:1990dx} and 
constraints from $b$ physics experiments~\cite{Amhis:2012bh}. In particular, we take into 
account constraints from $b\rightarrow s\gamma$, $b\rightarrow s\mu^+\mu^-$ and $B_u\rightarrow\tau\nu$. 
We apply a lower limit on the chargino mass of 
$100$~GeV~\cite{Abbiendi:2002vz,Abdallah:2003xe,Abbiendi:2003sc}. Finally, we demand a 
sufficiently large mass splitting, $m_{\tilde{\chi}^\pm_1} - m_{\tilde{\chi}^0_1} > \mev{280}$, between the higgsino-like NLSP ($\tilde\chi_2^0,\tilde\chi_1^\pm$) and the LSP ($\tilde\chi_1^0$), allowing for a prompt decay, so we do not consider searches for long-lived SUSY particles.

Since $R$-parity is conserved, the LSP is stable and  contributes to the dark matter relic density. For 
standard cosmology scenarios, the abundance is determined by the thermally averaged annihilation 
cross section, and the resulting relic density has to satisfy the precision measurement of the 
dark matter density from the cosmic microwave background. However, a higgsino dark matter particle
efficiently annihilates into SM particles and satisfies the upper limit on the relic density unless 
$m_{\tilde{\chi}^0_1}\gsim1$~TeV \cite{Bramante:2014tba}, which is satisfied by all of our points. In addition, if we extended our minimal natural SUSY scenario with a bino-like
neutralino LSP or a singlino LSP, the observed dark matter abundance could be explained in a large region of our parameter space \cite{Boehm:1999bj,Kim:2014noa}.

\begin{figure} 
\includegraphics[width=0.5\textwidth]{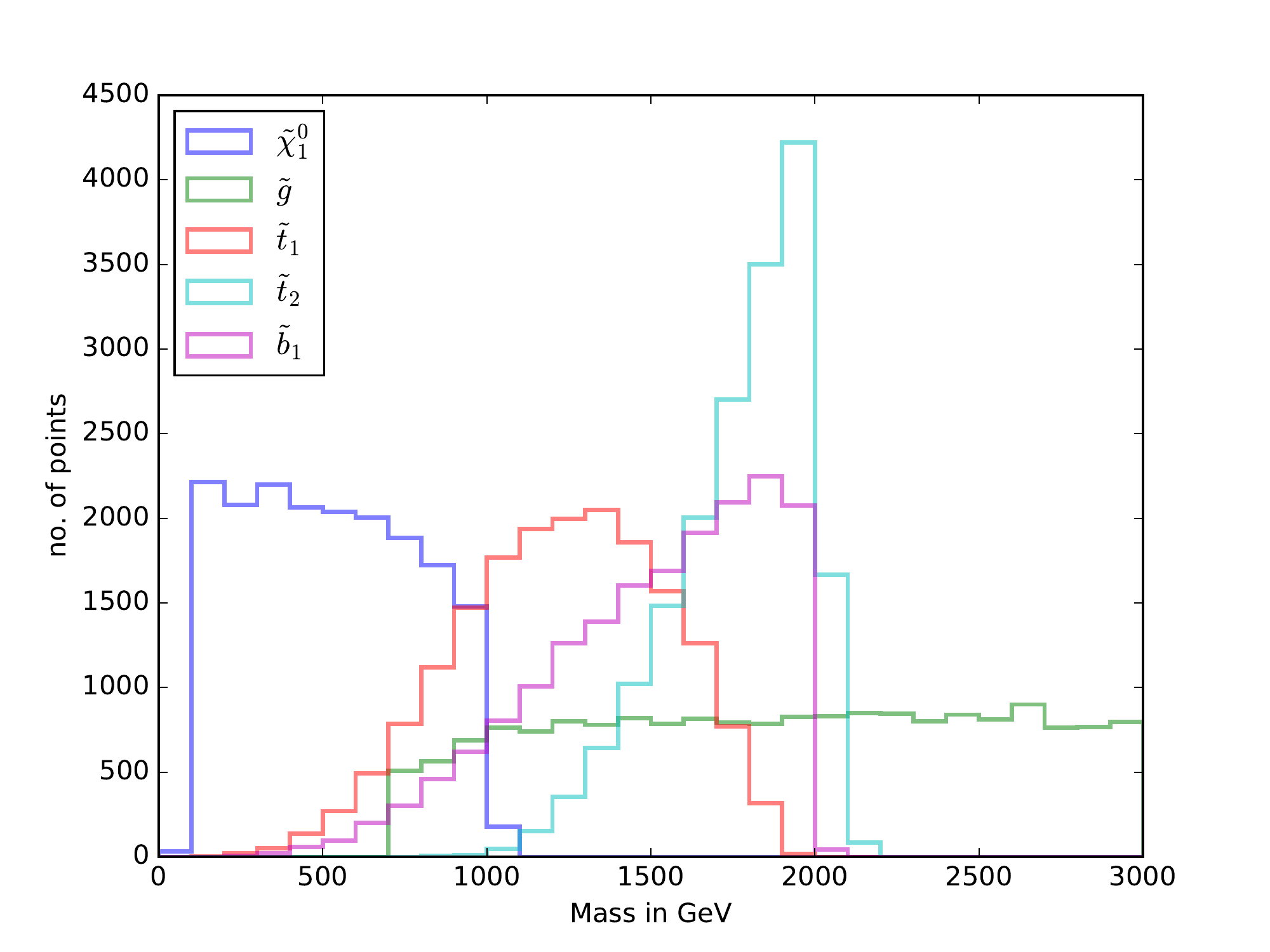} 
\caption{Distributions for the masses of the SUSY particles, $\tilde{\chi}^0_1$, $\tilde{g}$,
$\tilde t_1$, $\tilde t_2$ and $\tilde b_1$, for 
our model points satisfying the preselection criteria.}
\label{fig:sparticle_mass} 
\end{figure}

From the 150000 benchmark points initially generated, 20000 pass all preselection 
cuts. We 
show in Fig.~\ref{fig:sparticle_mass} the distributions of the higgsino, third-generation scalar quark and gluino masses. 
The higgsino is the lightest sparticle in 
our spectrum and has a reasonably flat mass distribution. However, it decreases at its 
upper limit since in this region of parameter space the higgsino is quite often 
not the LSP, consequently, these benchmark points are removed. 

Only a few benchmark points contain a very light stop with masses below 500~GeV. A 
light $\tilde t_1$ usually requires a heavy $\tilde t_2$ in order to increase the 
Higgs mass sufficiently. This can be seen in the rather large mass 
splitting between the lighter and heavier stop mass eigenstates. The direct stop searches 
of Run 1 are not sensitive to $\tilde t_2$ production for our parameter points 
since the cross section is too small, although indirect production via gluino decay 
may still be a non-negligible production channel. However, for Run 2 and for larger 
integrated luminosities, direct $\tilde t_2$ pair production may become accessible at the LHC. 

For the sbottom mass eigenstate $\tilde b_1$, the distribution covers a large range of 
masses. An SU(2)-doublet stop is accompanied by a sbottom with similar mass, and thus we expect for light left-handed stops, light sbottoms in the mass spectrum as well. The gluino mass distribution is flat above 1 TeV. It quickly falls off for gluino masses below 1 TeV since here the gluino could become lighter than the higgsino; hence these points would be removed.

Table~\ref{tbl:analyses} shows all relevant searches currently implemented in {\tt CheckMATE} and the 
left column provides the {\tt CheckMATE} identifier for each search. The second column 
gives the centre of mass energy and the third column displays the integrated luminosity that the search
has originally been tuned for, and the fourth column gives the target final state for which the 
corresponding search has been optimised for. The 
majority of the implemented searches are official ATLAS high luminosity studies which already 
cover a large number of our final state topologies. However, since we expect a 
large $b$-jet multiplicity in our natural SUSY scenarios, we have also included 
the current multi-$b$-jet and missing transverse momentum search 
at $\sqrt{s}=13$ TeV with a total integrated luminosity of $3.3$ fb$^{-1}$~\cite{ATLAS-CONF-2015-067}. 
Detailed information about the implemented searches can be found in Sec.~\ref{sec:experimental_analyses}.

The calculation procedure starts with the generation of Monte Carlo events, 
including all hadronic production processes given in Eq.~(\ref{eq:production}). The relative 
frequency of MC events is weighted according to the NLO+NLL cross section of the 
respective production channel. The hadronic MC events are then passed 
to {\tt CheckMATE} which simulates the ATLAS detector response with the fast detector 
simulation {\tt Delphes} which has a modified detector tuning and an extended 
list of final state objects. In particular, realistic $b$-jet tagging efficiencies and 
mistagging efficiencies were implemented. This is essential to obtain reliable results 
since a large number of natural SUSY searches in Table~\ref{tbl:analyses} rely 
on $b$-tagging in order to isolate the signal from SM backgrounds. {\tt CheckMATE} further 
processes the reconstructed detector-level objects with its analysis module. 

All searches listed in Table~\ref{tbl:analyses} have been carefully implemented and 
validated, and detailed information on the validation can be found on the official \texttt{CheckMATE} 
web page~\cite{webpage}.
Each model point was tested against all the analyses given in 
Table~\ref{tbl:analyses}. To do this, {\tt CheckMATE} calculates the efficiencies for 
all signal regions of all employed searches, and the program then chooses the signal region with 
the most sensitivity. To determine whether the model point in question is excluded 
at the $95\%$ C.L.~\cite{Read:2002hq}, we evaluate the ratio
\begin{equation} \label{r}
r \equiv \frac{S-1.96\cdot\Delta S} {S_{\rm exp.}^{95}}\,, 
\end{equation}
where $S$ is the number of expected signal events, $\Delta S$ denotes its theoretical uncertainty, 
and $S_{\rm exp.}^{95}$ is the theoretically determined 95$\%$ C.L. limit on the signal 
depending on the number of background events and its statistical as well as systematic 
error. {\tt CheckMATE} does not statistically combine signal regions of all the 
employed ATLAS searches. We consider a model to be excluded at 95$\%$ C.L.\  if $r$ 
defined in Eq.~(\ref{r}) exceeds $1$. 

In addition, we also display points that are confidently allowed at a certain energy and luminosity.
Here we use an adjusted $r'$ value, given as
\begin{equation} \label{r_prime}
r' \equiv \frac{S+1.96\cdot\Delta S} {S_{\rm exp.}^{95}}\,, 
\end{equation}
and define the point as allowed if $r'<1$. Finally, we define points as ``ambiguous'' if neither
of the above conditions are met.

In this work we are also interested in the discovery potential, and for this calculation we adopt 
the method called $Z_{Bi}$ in Ref.~\cite{Linnemann:2003vw} due to its optimality properties. 
The algorithm incorporates uncertainties in the background estimation in a fully frequentist fashion.   

\begin{table*}[t]
		\begin{tabular}{lccllr} \hline\hline
			Analysis & Energy\;\; & Luminosity\;\;  & Process\;\; & Final state & Reference \\ 
			(\texttt{CheckMATE} identifier)         & (TeV)  & (fb$^{-1}$) & targeted\;\; &  &  \\ \hline
			\texttt{atlas\_conf\_2015\_067} & 13 & 3.3 & $\tilde g \tilde g$ & $\geq$ 3 $b$-jets + MET & \cite{ATLAS-CONF-2015-067} \\
			\texttt{atlas\_phys\_2014\_010\_300} & 14 & 300 & $\tilde g \tilde g$, $\tilde q \tilde q$ & 2-6 jets + MET & \cite{ATL-PHYS-PUB-2014-010} \\
			\texttt{atlas\_phys\_2014\_010\_hl\_3l} & 14 & 3000 & $\tilde{\chi}_1^\pm \tilde{\chi}_2^0$ & 3 leptons + MET & \cite{ATL-PHYS-PUB-2014-010} \\
			\texttt{atlas\_phys\_2014\_010\_sq\_hl} & 14 & 3000 & $\tilde g \tilde g$, $\tilde q \tilde q$ & 2-6 jets + MET & \cite{ATL-PHYS-PUB-2014-010} \\
			\texttt{atlas\_phys\_pub\_2013\_011} & 14 & 3000 & $\tilde t \tilde t$ & 0-1 lepton + $\geq$ 3 jets + $\geq$ 1 $b$-jet + MET & \cite{ATL-PHYS-PUB-2013-011} \\
			\texttt{atl\_phys\_pub\_2014\_010\_sbottom} & 14 & 300 & $\tilde b \tilde b$ & 2 $b$-jets + MET & \cite{ATL-PHYS-PUB-2014-010} \\
			\hline\hline
		\end{tabular}
		\caption{Summary of the analyses used to test models, along with the energy and luminosity that
		they have originally been optimised for.}
\label{tbl:analyses}
\end{table*}	

\subsection{Rescaling due to energy and luminosity}\label{subsec:rescaling}
To be able to investigate all the combinations of centre-of-mass energy and integrated luminosity described above, 
a rescaling procedure has been used. We use a rescaling since there are currently only a handful of 
searches published by the ATLAS and CMS collaborations that target relevant processes at the high luminosity LHC.
After trying to fully reproduce all significant standard model and detector backgrounds, we found
that such a rescaling is more accurate than, for example, trying to reproduce the correct multi-jet background
estimation using a fast detector simulation.

All the searches used are listed in Table~\ref{tbl:analyses}. We can see that most analyses are for \tev{14} and $\ifb{3000}$ and 
no analyses are available for \tev{13} and \ifb{20} or \ifb{100}.
Since we want to compare the exclusion reach of the LHC at these various steps, we rescale the background numbers given by the 
original searches to the centre-of-mass energies and integrated luminosities that we need.
We also rescale the generated signal events in the same way to correctly reproduce the signal-to-background ratio at different energies. 
The exact rescaling procedures are described below.

\subsubsection{Rescaling the searches}
We start the discussion with rescaling the searches to the different integrated luminosities and centre-of-mass energies considered.
This method will later be used in \texttt{CheckMATE} to compute new $S^{95}_{\rm exp}$ values that are used to test model points.
The calculation requires the number of expected background events and its uncertainty for each signal region.
Two different rescaling procedures are used: one to change the centre-of-mass energy and the other to change the integrated luminosity.

\paragraph{Changing the centre of mass energy}
To change the centre-of-mass energy of a search, the number of background events and their uncertainties are rescaled with the ratio of the tree-level cross sections at the two centre-of-mass energies.
The cross sections used for this rescaling were obtained using \texttt{MadGraph} \cite{Alwall:2011uj,Alwall:2014hca}.
As a consequence, we assume that the ratio of the signal-to-background kinematic distributions will not change significantly between \tev{13} and \tev{14}, or at least 
similarly enough so that the signal regions defined by the searches are insensitive to the difference within errors.

To rescale the background numbers, every background process is scaled separately, and then all processes are summed 
to give the total number of background events at the new centre-of-mass energy.
For some of the searches, background numbers are only listed for the most significant background processes.
In that case the sum of the individual numbers of background events is not equal to the total number of background events 
provided by the ATLAS note.
When rescaling such a search, the relative difference between the total background and the sum of the 
individual processes is kept constant.
The uncertainties on the backgrounds are scaled with the same factor as the corresponding background 
and are then combined in the same way as in the original study to give the total background uncertainty.
Due to some of the searches missing background contributions and uncertainty correlations being taken
into account, a simple combination is not always possible.
To estimate these effects, we follow the same approach as for the total numbers and keep the relative difference between the 
total and the sum of the individual contributions constant.

\paragraph{Changing the integrated luminosity}
When changing the integrated luminosity from $l_1$ to $l_2$, the number of background events and the uncertainty are 
scaled with the ratio $l_2/l_1$.
This assumes that nothing other than the number of expected events (signal or background) changes between 
the two luminosities, and for example, possible detector upgrades or differences in pileup are not taken into account.

\subsubsection{Rescaling the signal events}
When \texttt{CheckMATE} tests a parameter point against a particular analysis, it sums the weights of the 
generated signal events that fall into each signal region. 
To compare this number against the expected number of background events, \texttt{CheckMATE} normalises the weighted MC events 
to the cross section of the process and the integrated luminosity for the analysis.
When rescaling the signal events, the same normalisation procedure is repeated with the 
desired integrated luminosity and with the cross section of the process at the desired centre-of-mass energy.
As with the rescaled backgrounds, the effects of different acceptance times efficiency following 
from modified kinematical distributions at different centre-of-mass energies are neglected.
Also, as for the backgrounds, different pileup and possible detector improvements for 
the various integrated luminosities are not taken into account.
The uncertainties on both the number of signal events and on the cross section are normalised 
using the same factor as for the signal events.
Consequently, the relative uncertainties remain unchanged after rescaling.

\subsubsection{Reduced systematic uncertainty scenario\label{sec:reducedsyst}}
To account for possible future improvements in the various systematic uncertainties, we also consider a scenario 
where these uncertainties are reduced with the integrated luminosity.
In this case we fix the values of errors at \ifb{20} and then assume that the systematic uncertainties 
scale like a statistical error with added data; i.e., the relative uncertainty decreases with the square root of 
luminosity, $\propto 1/\sqrt{\mathcal{L}}$.
This procedure is performed for the uncertainty on the number of background events and on the cross section.
These uncertainties can be reasonably expected to improve in the coming years with a better 
understanding of the detectors resulting in reduced systematics and, from the theory side, with more 
precise calculations of the relevant background and signal processes.
Additional improvements are expected for the parton distribution functions at the 
higher centre-of-mass energies probed now at the LHC.

Nevertheless, it is extremely hard to predict how these uncertainties will reduce in the coming years.
We choose to reduce the errors as a statistical uncertainty since the dominant LHC paradigm is to normalise
backgrounds with dedicated control regions and only use theoretical predictions to extrapolate these results
into signal regions. Obviously, these extrapolations are independent of the collected statistics but can be improved
beyond the naive theoretical uncertainty by combining the results of many different independent validation and
control regions (especially for processes like $t\bar{t}$ production with many different final states to study).
Any estimate is further complicated by the fact that there is a different proportion of background final
states in each individual signal region we examine. In addition, each signal region probes a different amount
into the tail of the corresponding background distribution.

In conclusion, to make a more thorough prediction of the background uncertainty would require a detailed analysis
of each individual signal region. This would entail a significant amount of work but would still be extremely 
speculative in nature. Therefore we believe that the far simpler and transparent procedure of simply 
reducing the uncertainty due to statistics is more sensible. This should be considered as the ``most'' optimistic
scenario, and our opinion is that reality will lie somewhere between the two extremes presented.

Explicitly, the method we use is that the systematic uncertainties are first scaled 
linearly down to $l_b$ from $l_1$, where $l_b$ is the base luminosity of \ifb{20} 
and $l_1$ the original luminosity for a given search.
The uncertainties are then scaled up to the new luminosity $l_2$ with $\sqrt{l_2/l_b}$, giving
the complete expression as
\begin{equation}
	\sigma'(l_2) = \frac{l_b}{l_1}\sqrt{\frac{l_2}{l_b}}\; \sigma(l_1),
\end{equation}
where $\sigma$ $(\sigma')$ is either the original (rescaled) uncertainty on the number 
of background events or the uncertainty of the cross section. The above expression can be trivially
rewritten as
\begin{equation}
	\frac{\sigma'(l_2)}{l_2} = \sqrt{\frac{l_b}{l_2}}\; \frac{\sigma(l_1)}{l_1}\;,
\end{equation}
which explicitly shows that the relative uncertainty scales as $1/\sqrt{\mathcal{L}}$.

\subsection{Experimental analyses considered\label{sec:experimental_analyses}  }  

The backbone of our study are the searches ATLAS proposed for HL-LHC analyses \cite{ATL-PHYS-PUB-2014-010}. These analyses 
take into account the upgraded ATLAS detector configuration, HL-optimised selections and MC-derived 
estimation of background processes. They target the following SUSY benchmark processes:
\begin{itemize}
 \item direct production of charginos and neutralinos with decays via $W$, $Z$, $h$;
 \item production of squarks (1st and 2nd generation) and gluinos;
 \item production of bottom squarks.
\end{itemize}
The assumed centre-of-mass energy is 14~TeV, and the integrated luminosities 
of $300~\mathrm{fb}^{-1}$ and $3000~\mathrm{fb}^{-1}$ were considered. The searches have now
been incorporated into {\tt CheckMATE} and are summarised in Table~\ref{tbl:analyses}. Our 
validation was performed based on the information provided in Ref.~\cite{ATL-PHYS-PUB-2014-010}, 
both for signal and background processes.    

Since stop production is also clearly important in our natural SUSY scenario, we also include
the separate high luminosity stop analysis \cite{ATL-PHYS-PUB-2013-011}. This is also optimised
for $\sqrt{s} = \tev{14}$, with a final state containing several jets, 
at least one $b$-jet, 0--1 leptons and missing transverse energy; see Table~\ref{tbl:analyses}.

We also include a recent study searching for the direct production of gluinos, based on 
data collected during the first phase of Run 2 in 2015. The study targets 
multi-$b$-jets final states with additional missing transverse energy. While it was 
optimised for low integrated luminosity, it may still be relevant for gluino 
exclusion when we consider prospects at $20~\mathrm{fb}^{-1}$ and relatively light 
gluinos. We note here that the high luminosity analyses optimised for heavy SUSY particles 
could have a limited sensitivity for gluinos just above the current exclusion limits because the
signal regions are tuned for higher mass states.  

Finally, we also retest points that cannot be excluded with the above searches, with the full
suite of \texttt{CheckMATE} analyses at 8~TeV. As will be seen in the results section, there are
a few points containing light stops that the high energy analyses are insensitive to because 
they are tuned to higher mass SUSY production. However, these points are already excluded
by 8~TeV searches \cite{Aad:2015zva, Aad:2013ija, Aad:2014wea, Aad:2014lra}.

\section{Results}
\label{sec:results}

As described in the numerical analysis section, we study four different collider scenarios
which reflect the expected progress of the LHC in the coming years. Namely, we
consider 20~fb$^{-1}$ and 100~fb$^{-1}$ at 13~TeV, which is the current integrated luminosity 
estimate for Run 2 before the long shutdown begins in late 2018. We make the assumption
that by 2021 the full capability of the LHC will be achieved, and the restart will
be at 14~TeV. Here we study two integrated luminosity scenarios, 300~fb$^{-1}$ and
3000~fb$^{-1}$.

\subsection{Exclusion with 20~fb$^{-1}$ at 13~TeV}

\begin{figure*}[ht] 
\includegraphics[width=\textwidth]{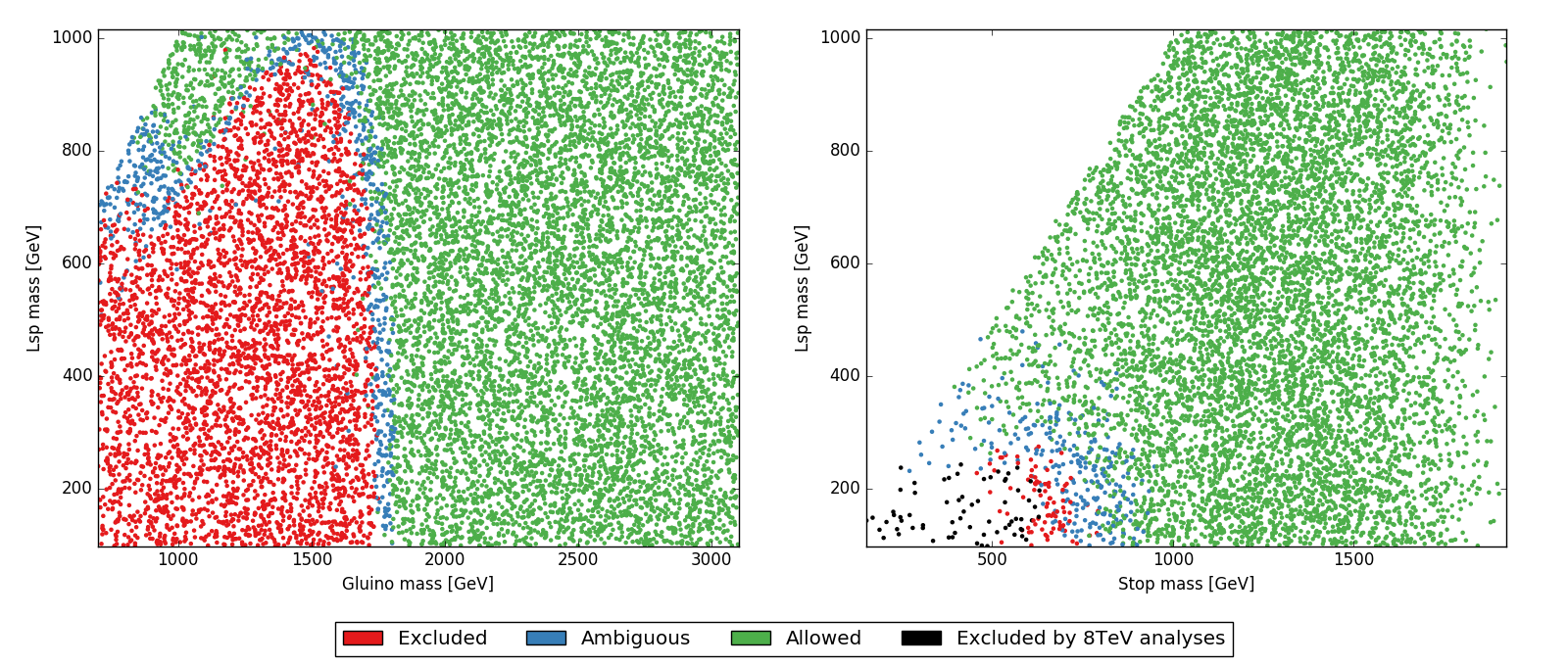} 
\caption{Plots showing the natural SUSY points that are allowed, excluded or ambiguous within the Monte Carlo
uncertainty at $\sqrt{s} = \tev{13}$ with $\mathcal{L} = \ifb{20}$, under the assumption that the current systematic
errors will remain constant. 
Left: $m_{\tilde{g}}$ vs $m_{\tilde{\chi}^0_1}$ for $m_{\tilde{t}_1}>1000$~GeV. 
Right: $m_{\tilde{t}_1}$ vs $m_{\tilde{\chi}^0_1}$ for $m_{\tilde{g}}>2000$~GeV.
}
\label{fig:exclu_20_13}  
\end{figure*} 

To begin, we examine our baseline study of 20~fb$^{-1}$ data collected at 13~TeV. If we
look at Fig.~\ref{fig:exclu_20_13} (left), we plot the excluded,
ambiguous or allowed points in the $m_{\tilde{g}}$ vs $m_{\tilde{\chi}^0_1}$ plane
in scenarios with $m_{\tilde{t}_1}>1000$~GeV. We remove scenarios with
light $\tilde{t}_1$ states so that we can clearly see the dependence on
the $\tilde{g}$ mass. Once these points have been removed,
we can very clearly see that 
the phenomenology of the model in this region of parameter space
is dominated by the gluino and LSP masses. More precisely, we see that 
scenarios with gluino masses up to 1700~GeV can reliably be excluded if
the parameter point contains relatively light LSP masses ($m_{\tilde{\chi}^0_1}<600$~GeV). Once
the LSP mass becomes larger, however, for the compressed part of the 
spectra, $m_{\tilde{g}} - m_{\tilde{\chi}^0_1} \lsim \gev{300}$, 
the exclusion is no longer reliable. The lack of exclusion in compressed
spectra is a well-known phenomenon that occurs because the decay
products of the gluino become much softer and are thus harder to
distinguish from the SM background \cite{LeCompte:2011cn,Dreiner:2012gx,Dreiner:2012sh}. The largest excluded LSP
masses we find have $m_{\tilde{\chi}_1^0} \sim \gev{1000}$ and are reached when $m_{\tilde{g}} \simeq \gev{1500}$.   

We now move on to concentrate on the light $\tilde{t}_1$ scenarios that we have just projected out.
To do this we plot in Fig.~\ref{fig:exclu_20_13} (right) the excluded,
ambiguous or allowed points in the $m_{\tilde{t}_1}$ vs $m_{\tilde{\chi}^0_1}$ plane
in scenarios with $m_{\tilde{g}}>2000$~GeV. The light gluino parameter points are
removed for exactly the same reasoning as above so that we can concentrate on the
light $\tilde{t}_1$ phenomenology. We find that for a light LSP 
($m_{\tilde{\chi}^0_1} < 200$~GeV), stops can be excluded up to $m_{\tilde{t}_1}<800$~GeV.

As for the gluino limits, we find that as the LSP becomes heavier ($m_{\tilde{\chi}^0_1} > 200$~GeV), the
stop limits that we can set, quickly degrade. Again, the reason is that as the scenarios become compressed,
the energy of the visible final state is significantly reduced and the analyses lose sensitivity \cite{Carena:2008mj,Drees:2012dd,Bornhauser:2010mw}.
As an extreme
example, we find points with $m_{\tilde{t}_1}<500$~GeV still allowed when $m_{\tilde{\chi}^0_1} \sim 300$~GeV. However, we want to point out 
that very long-lived stops could be observed at the LHC via $R$ hadron searches \cite{Hiller:2009ii,Aad:2013gva}.
We also note that we found a large cluster of points at low stop mass that could not be excluded by
the high energy searches that we used. However, the reason is not that these points cannot be excluded
at 13~TeV with 20~fb$^{-1}$ but rather that the searches we use have signal regions tuned for higher mass states.
To illustrate our point, we retest all of the light points that cannot be excluded with the 8~TeV searches
implemented in \texttt{CheckMATE} and these are given in black. Therefore, we see that the 8~TeV signal regions
that have been tuned better for these stop masses successfully cover the low mass region.

\subsection{Exclusion with 100~fb$^{-1}$ at 13~TeV}

\begin{figure*} 
\includegraphics[width=\textwidth]{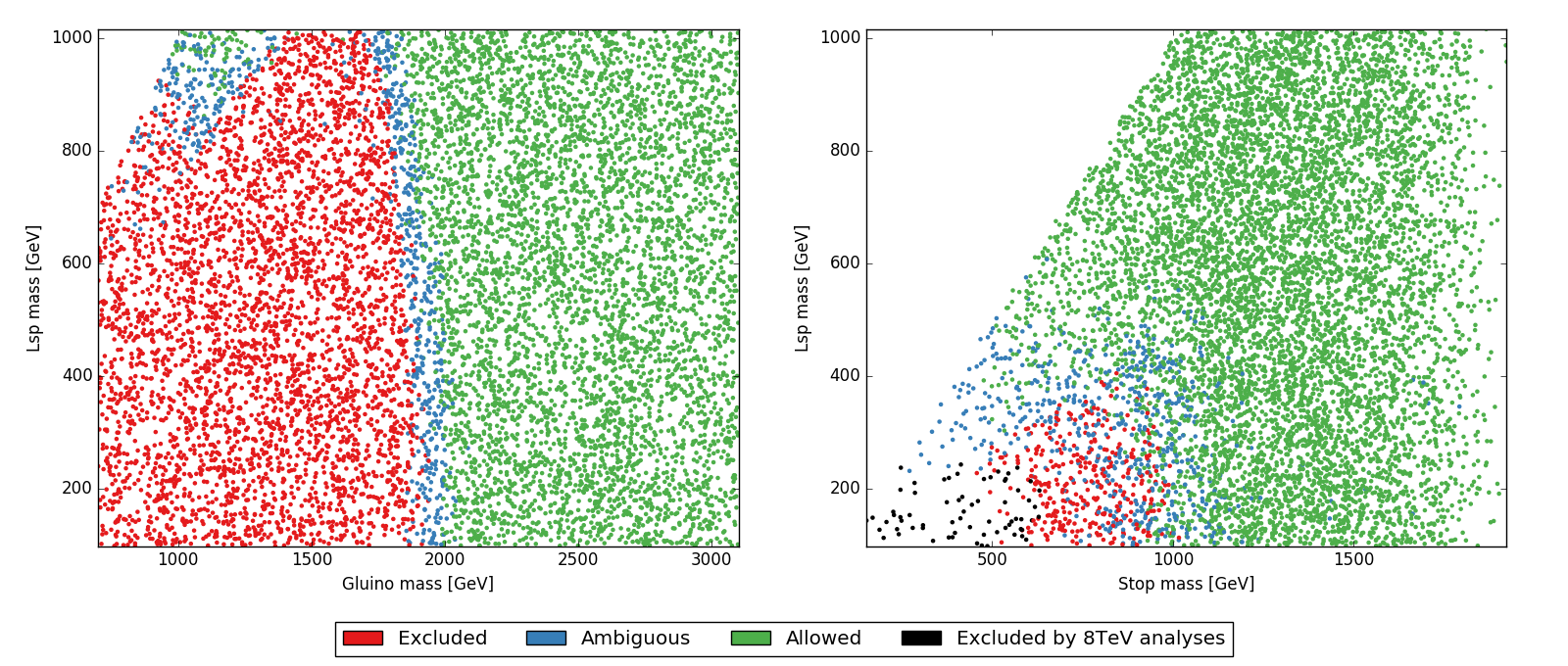} 
\caption{Plots showing the natural SUSY points that are allowed, excluded or ambiguous within the Monte Carlo
uncertainty at $\sqrt{s} = \tev{13}$ with $\mathcal{L} = \ifb{100}$, under the assumption that the current systematic
errors will remain constant. 
Left: $m_{\tilde{g}}$ vs $m_{\tilde{\chi}^0_1}$ for $m_{\tilde{t}_1}>1200$~GeV. Right: $m_{\tilde{t}_1}$ vs $m_{\tilde{\chi}^0_1}$ for $m_{\tilde{g}}>2000$~GeV.
}
\label{fig:exclu_100_13}  
\end{figure*} 

If we stay at 13~TeV but now move to 100~fb$^{-1}$, we see that the exclusion bounds for both
the gluino and stop increase but only by a small amount. More precisely, Fig.~\ref{fig:exclu_100_13} 
shows that the exclusion on the gluino mass now reaches almost 2000~GeV while $m_{\tilde{t}_1} > \gev{1200}$. More important, however, is 
the observation that the number of compressed points that are definitely allowed is 
significantly reduced. If we make the assumption that searches will exist that target this region
with specifically tuned cuts, the compressed region can be excluded up to $m_{\tilde{\chi}_1^0} \simeq \gev{900}$.

Moving on to the stop searches, we see a similar improvement. Parameter points with $m_{\tilde{t}_1} < 1000$~GeV
are now regularly excluded, but we should also make it clear that some points are allowed even though they 
have $m_{\tilde{t}_1}\sim 900$~GeV and a light LSP with $m_{\tilde{\chi}^0_1}\sim 200$~GeV. Once again,
we see a reduction in the limit as the LSP mass is increased. However, the light stop points 
($\tilde{t}_1<500$~GeV) that were previously allowed with 20~fb$^{-1}$ are now in the ``ambiguous''
region and could probably be excluded with targeted searches.

\subsection{Exclusion with 300~fb$^{-1}$ at 14~TeV }

\begin{figure*} 
\includegraphics[width=\textwidth]{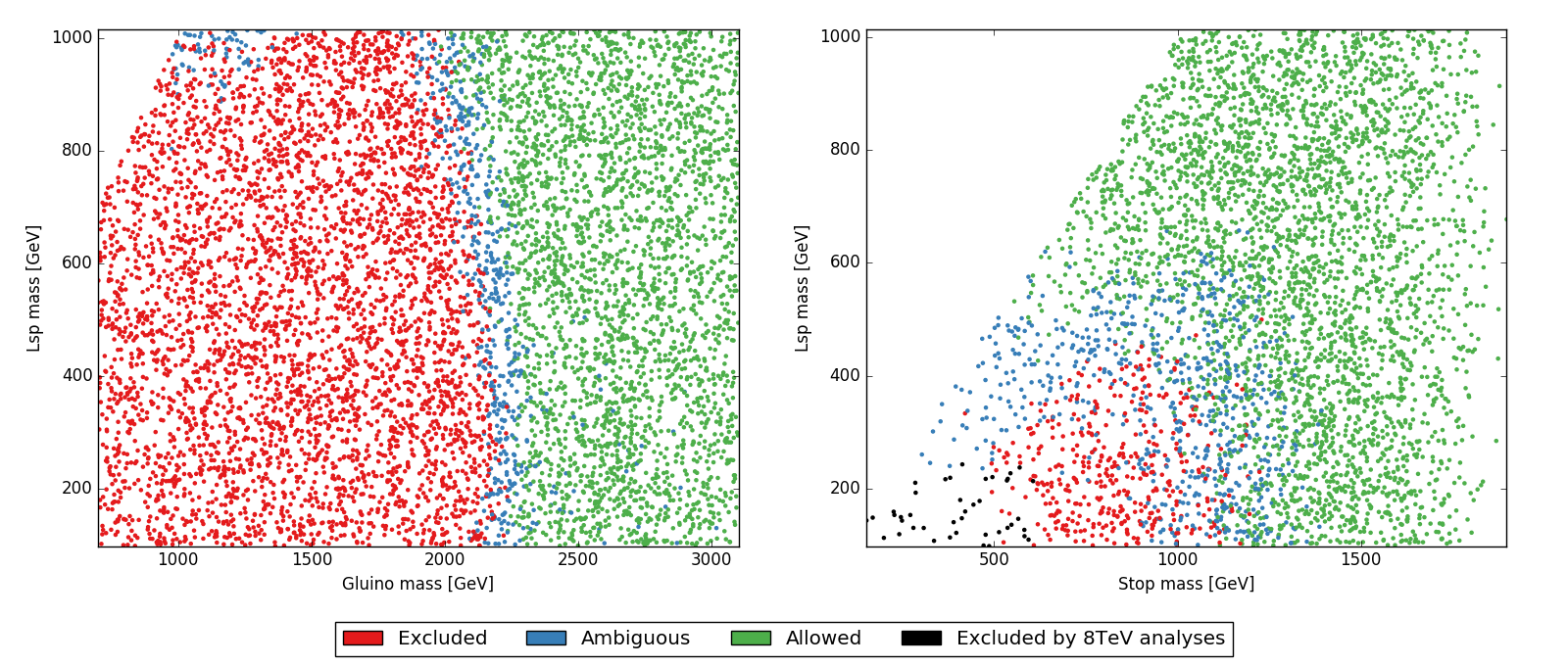}
\caption{Plots showing the natural SUSY points that are allowed, excluded or ambiguous within the Monte Carlo
uncertainty at $\sqrt{s} = \tev{14}$ with $\mathcal{L} = \ifb{300}$, under the assumption that the current systematic
errors will remain constant. 
Left: $m_{\tilde{g}}$ vs $m_{\tilde{\chi}^0_1}$ for $m_{\tilde{t}_1}>1300$~GeV. Right: $m_{\tilde{t}_1}$ vs $m_{\tilde{\chi}^0_1}$ for $m_{\tilde{g}}>2500$~GeV.
}
\label{fig:exclu_300_14}  
\end{figure*} 

In Fig.~\ref{fig:exclu_300_14} we display the points predicted for exclusion with 300~fb$^{-1}$
at 14~TeV, and as expected, we see that the bounds further improve from those obtained at 13~TeV. For
gluinos, assuming $m_{\tilde{t}_1} > \gev{1300}$, Fig.~\ref{fig:exclu_300_14} (left), the exclusion in some parameter points extends to 
over $m_{\tilde{g}}\sim2200$~GeV, and
the exclusion is almost complete in the compressed region even as the LSP mass reaches 
$m_{\tilde{\chi}^0_1}\sim1000$~GeV. 

In order to isolate the dependence of the LHC reach on
the stop mass, we now have to place a far harder cut on the gluino 
mass ($m_{\tilde{g}}>2500$~GeV) in our selection, and this leads to the 
sparser coverage of points shown in Fig.~\ref{fig:exclu_300_14} (right). However,
the plot still clearly shows the improved sensitivity to stop states, with masses
above $m_{\tilde{t}_1}\sim 1200$~GeV now probed for direct production.
Nonetheless, the coverage is not complete and much lighter stop 
states ($m_{\tilde{t}_1}<1000$~GeV) are in the ``ambiguous'' category even with 
an LSP of $m_{\tilde{\chi}^0_1}\sim100$~GeV. 

\subsection{Exclusion with 3000~fb$^{-1}$ at 14~TeV }

\begin{figure*} 
\includegraphics[width=\textwidth]{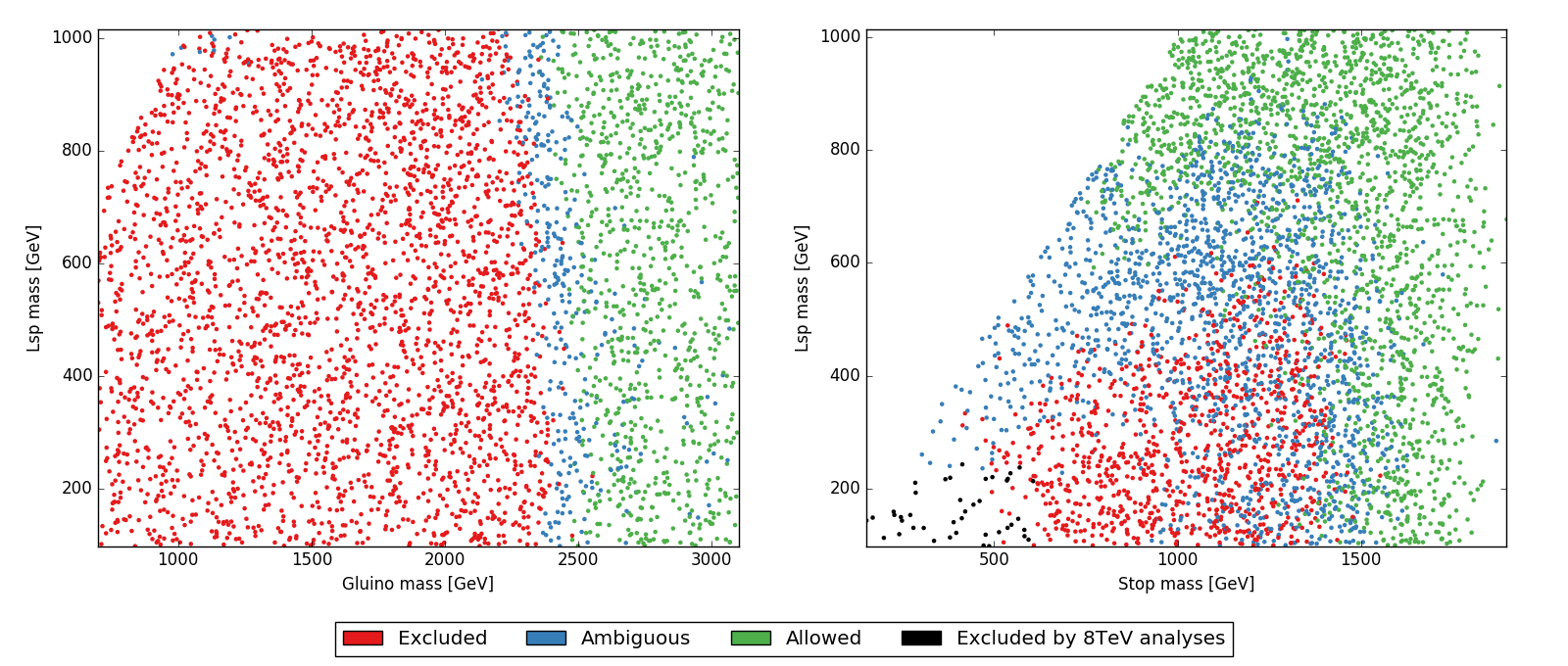}
\caption{Plots showing the natural SUSY points that are allowed, excluded or ambiguous within the Monte Carlo
uncertainty at $\sqrt{s} = \tev{14}$ with $\mathcal{L} = \ifb{3000}$, under the assumption that the current systematic
errors will remain constant. 
Left: $m_{\tilde{g}}$ vs $m_{\tilde{\chi}^0_1}$ for $m_{\tilde{t}_1}>1500$~GeV. Right: $m_{\tilde{t}_1}$ vs $m_{\tilde{\chi}^0_1}$ for $m_{\tilde{g}}>2500$~GeV.
}
\label{fig:exclu_3000_14}  
\end{figure*} 

Further evolution in the bounds is seen as we move to the final luminosity estimates
expected for the HL LHC, and these are shown in Fig.~\ref{fig:exclu_3000_14}.
We see that gluinos can now be excluded up to almost $m_{\tilde{g}}\sim2500$~GeV for certain 
scenarios, and all points with $m_{\tilde{g}} < 2000$~GeV can definitely be excluded as long
as the parameter points are not highly compressed.
Even in the compressed region, however, we see very good
coverage with almost all points at $m_{\tilde{\chi}^0_1}\sim m_{\tilde{g}}\sim1000$~GeV
completely excluded and only a few remaining in the ambiguous category.

For stops the limits can now even reach beyond $m_{\tilde{t}_1}\sim 1500$~GeV for the most sensitive 
parameter points. However, we still see significant variation across the parameter space, 
and some models with a light stop with $m_{\tilde{t}_1}< 1000$~GeV are in the ambiguous
category, which we cannot be sure will be excluded. Some of this variation is due to differences
in how the stop states can decay but the major difference is down to the masses of the other 
coloured scalars in the theory.

For example, if we examine the ambiguous points in Fig.~\ref{fig:exclu_3000_14} (right) 
with $m_{\tilde{t}_1}< 1000$~GeV and $m_{\tilde{\chi}^0_1}\sim 100$~GeV, we find that 
the light stop in these parameter points is the right-handed SU(2) singlet. It decays at comparable rates to different final states 
(see Eqs.~\eqref{eq:stopdec1} and \eqref{eq:stopdec2} and the following discussion), which weakens sensitivity of the individual searches. The SU(2)-doublet 
stop (along with the partner sbottom) by contrast is effectively decoupled
from the LHC phenomenology with masses ${\sim} 2000$~GeV. In addition, the gluinos are very
heavy for these points, with $m_{\tilde{g}}\sim 3000$~GeV.

A comparison to these light stop points is provided by a spectra where the lightest stop
has $m_{\tilde{t}_1}\sim 1500$~GeV but the point can still be excluded. The reason is that
all of the coloured scalars ($\tilde{t}_1, \tilde{t}_2, \tilde{b}_1$) in the model
have a very similar mass ($\sim 1500$~GeV), and this provides a significant production
cross section. In addition, while the gluino is not light enough by itself to result
in the parameter point being excluded, the mass ($m_{\tilde{g}}\sim2560$~GeV) is in the borderline region, where 
the production cross section still provides additional events for the dedicated stop searches.

\subsection{Exclusion with 3000~fb$^{-1}$ at 14~TeV: Reduced~errors}
\label{sec:exclu_3000_14_sqrtErr}

\begin{figure*} 
\includegraphics[width=\textwidth]{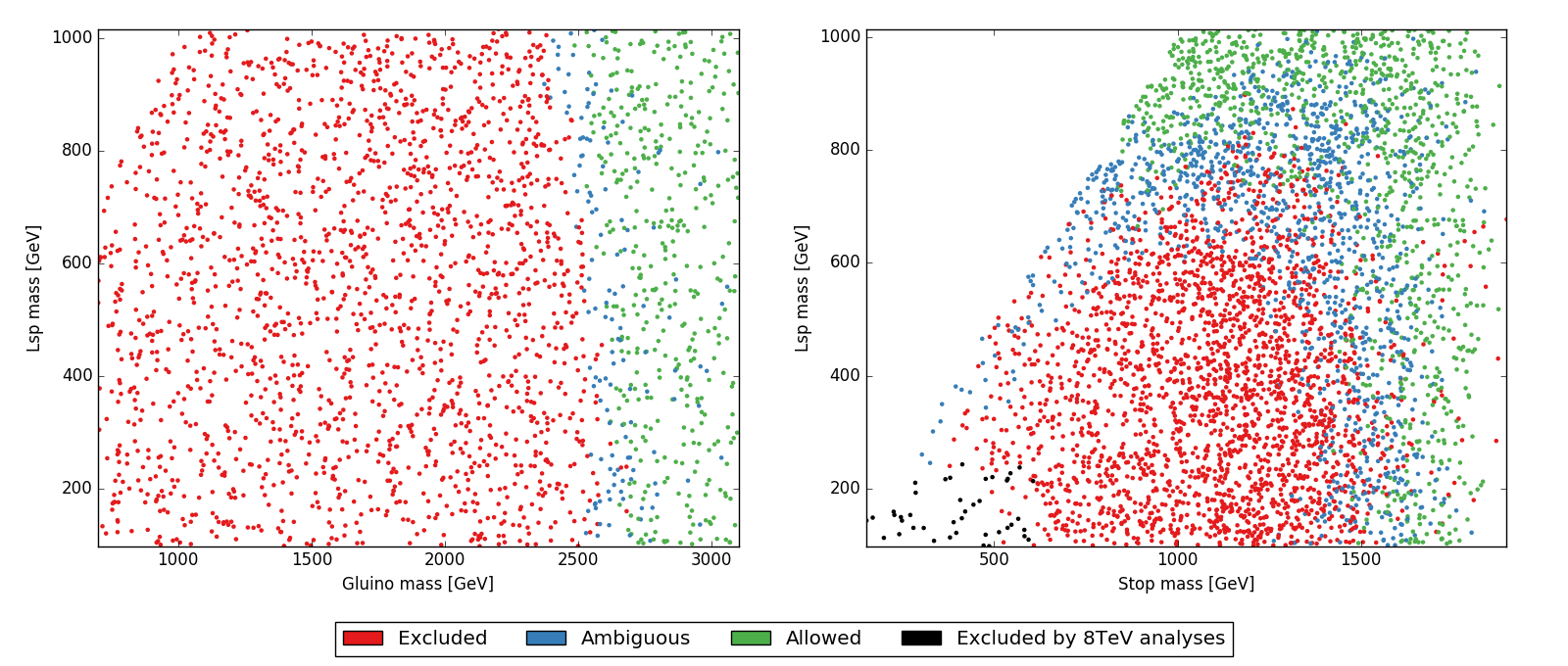} 

\caption{Plots showing the natural SUSY points that are allowed, excluded or ambiguous within the Monte Carlo
uncertainty at $\sqrt{s} = \tev{14}$ with $\mathcal{L} = \ifb{3000}$, under the assumption that the systematic
errors will reduce in proportion to the collected luminosity. 
Left: $m_{\tilde{g}}$ vs $m_{\tilde{\chi}^0_1}$ for $m_{\tilde{t}_1}>1600$~GeV. Right: $m_{\tilde{t}_1}$ 
vs $m_{\tilde{\chi}^0_1}$ for $m_{\tilde{g}}>2500$~GeV.
}
\label{fig:exclu_3000_14_sqrtErr}  
\end{figure*} 

In all of the above results we have assumed that the systematic uncertainties
are constant (as a proportion) for all the different luminosities and energies that
we investigate. However, it can be expected that as the LHC develops, these uncertainties
will reduce as the collider and detectors become better understood. In particular, many
backgrounds do not rely on MC predictions but are derived from data, and thus
the uncertainty can be expected to improve. As an approximation to 
illustrate how the reach of the LHC can improve as the uncertainties
are reduced, we take the current errors at 20~fb$^{-1}$ as the baseline and scale these
according to the collected statistics (i.e.\ the uncertainties reduce $\propto 1/\sqrt{\mathcal{L}}$,
see Sec.~\ref{sec:reducedsyst}).

Figure~\ref{fig:exclu_3000_14_sqrtErr} shows the expected parameter points that can be excluded, are
ambiguous or are expected to be allowed under the above assumption. We see that for the gluino limits
the effect is not so stark, and it simply increases the expected exclusion for almost all parameter
points investigated to $m_{\tilde{g}} > 2500$~GeV. 

The effect on the stop limits, however, is far stronger. Figure~\ref{fig:exclu_3000_14_sqrtErr} (right)
shows that the exclusion is possible for the majority of points with $m_{\tilde{t}_1} < 1400$~GeV and 
$m_{\tilde{\chi}_1^0} < \gev{600}$. 
This is a significant change compared to the scenario where the systematic errors remained unchanged
and some points with $m_{\tilde{t}_1}\sim 900$~GeV were still not definitely excluded. In addition, in
the region where the LSP is heavier and the scenario begins to become compressed the limits
also increase substantially. For example, without a reduction in the systematic error, no 
strict bound on the LSP mass for $m_{\tilde{t}_1}\sim1000$~GeV could be set. However, with
the reduced error we now find no parameter points that are allowed and only a few that are ambiguous for $m_{\tilde{t}_1}\sim1000$~GeV
with $m_{\tilde{\chi}^0_1}<600$~GeV.

One may ask, why the reach to stop states is improved far more with reduced systematic
uncertainties compared to the increase in the gluino limits. Our reasoning is that different
kinematics are probed by high-mass gluino production compared to the lower mass stop production.
Namely, the gluino searches are at the LHC kinematic limit when probing pair production with
$m_{\tilde{g}}\sim2500$~GeV. Here the parton distribution functions (PDFs) are dropping so quickly, even when we reduce the 
systematic errors significantly, that we simply lack enough signal events for the limits to increase
appreciably.

The stop limits are different since for their typical masses (${\sim} 1000$~GeV), we are much further away from the 
kinematic boundary where the PDFs rapidly reduce the production cross sections. Rather, the difficulty
in successfully setting limits is due to the large standard model background that comes predominantly
from top pair production. If we can successfully reduce the systematic error on this background,
far more of the parameter space can be probed.

\subsection{Most sensitive analyses}

\begin{figure*}[!ht] 
\includegraphics[width=\textwidth]{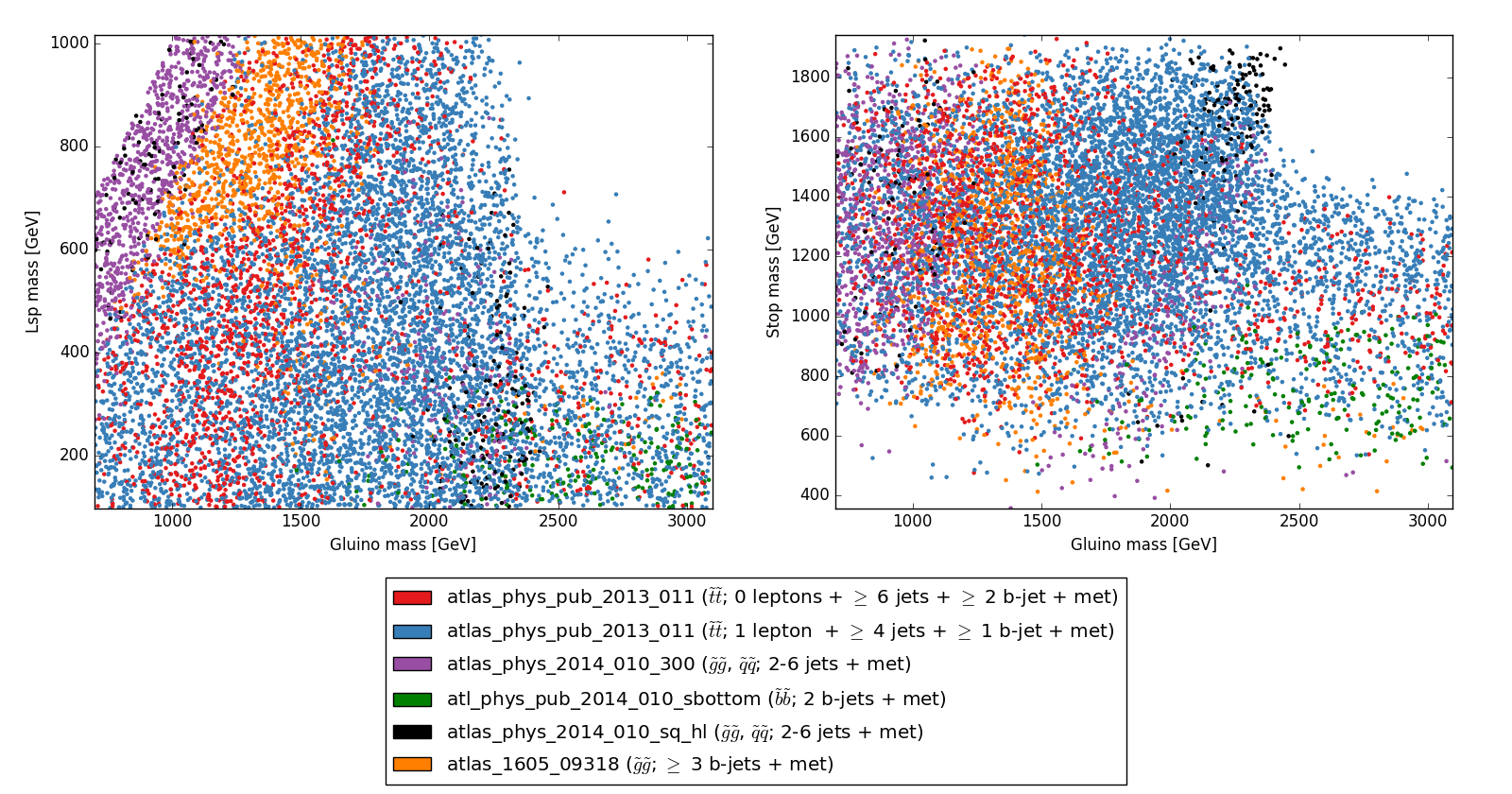} 
\caption{Most constraining analyses for each of the parameter points excluded at $\sqrt{s} = \tev{14}$ with $\mathcal{L} = \ifb{3000}$
assuming the systematic errors remain constant. Left: $m_{\tilde{g}}$ vs $m_{\tilde{\chi}^0_1}$.
Right: $m_{\tilde{g}}$ vs $m_{\tilde{t}_1}$. }
\label{fig:sensitive_analyses}  
\end{figure*} 

In Fig.~\ref{fig:sensitive_analyses} we plot the most sensitive analysis for every parameter point 
that can be excluded with 3000~fb$^{-1}$ at 14~TeV, assuming the systematic errors remain constant.
We can clearly see that the parameter space can be divided into regions where different searches
are most sensitive. If we first look at the $m_{\tilde{g}}$ vs $m_{\tilde{\chi}^0_1}$ plane 
(Fig.~\ref{fig:sensitive_analyses} (left)), we see that, in general, the most powerful search for kinematically
accessible gluinos ($m_{\tilde{g}}<2300$~GeV) and light LSPs ($m_{\tilde{\chi}^0_1}\lesssim600$~GeV)
is the one-lepton stop pair production search \cite{ATL-PHYS-PUB-2013-011} shown in blue. Since these points are dominated by gluino
production, this may, at first sight, seem surprising. However, the signal regions are reasonably general and 
rely on the following:
\begin{itemize}
 \item large missing energy, $E^{\text{miss}}_{\text{T}}$;
 \item transverse mass between the lepton and $E^{\text{miss}}_{\text{T}}$;
 \item $E^{\text{miss}}_{\text{T}}$ significance, $E^{\text{miss}}_{\text{T}}/\sqrt{H_{\text{T}}}$, where $H_{\text{T}}=\sum^4_{\mathrm{jets}=1}|\vec{p}_{\text{T}}|$; 
 \item reconstructed hadronic top, $130 < m_{jjj} < 205$~GeV.
\end{itemize}
Since the gluinos in our scenarios very commonly decay via cascades involving top quarks, it is easy to see why such
an analysis is so sensitive to our models. 

In similar regions of parameter space we also find that the 0-lepton version of the analysis~\cite{ATL-PHYS-PUB-2013-011} is sensitive (points in
red). Again, this is not surprising because the search concentrates on high missing energy along with high
jet multiplicity ($\geq 6)$, of which at least two must be $b$-jets. While these signal
regions are again labelled as stop searches we should point out
that they actually share more in common with the current gluino searches, and thus it is natural that they so
strongly constrain the gluinos in our model.

As we go to higher LSP masses ($m_{\tilde{\chi}^0_1}\gsim600$~GeV), we see a bulk region in orange that denotes
the parameter points expected to be excluded by the three $b$-jets search for gluinos \cite{ATLAS-CONF-2015-067}. The reason
why the general gluino search becomes more sensitive in this region is that the spectrum starts to become
more compressed as the LSP mass is raised. The consequence of such compression is that we produce fewer (or no) 
on-shell top quarks in the cascade decay chains. Consequently, the one-lepton stop search that requires a hadronically
reconstructed top quark is no longer so sensitive, and we rely on the simpler jet, $b$-jet and missing energy search.

To the left of the orange region is an area dominated by purple points that signify the general jets and missing
energy squark and gluino search but with the signal regions optimised for 300~fb$^{-1}$. This is the compressed 
region of parameter space where the gluino and LSP lie close in mass. Thus, the decay products of the gluino
become soft, and we rely on initial state radiation (ISR) in order to set limits on the model. Consequently, it is not 
so surprising that the dominant search in this region no longer relies on $b$-tags or leptons since these will,
in general, be softer and less likely to pass the kinematic cuts. In addition it is not surprising that the 300~fb$^{-1}$
version of the jets and $E^{\text{miss}}_{\text{T}}$ analysis performs better than the one optimised for 3000~fb$^{-1}$. 
The reason is that the cuts are softer for 300~fb$^{-1}$ signal regions, and thus the acceptance for the ISR signature
is much higher.

One may also ask how monojet 
searches may perform in this region since they can give the strongest bounds for very compressed models. However, this
is typically only the case for extreme compression ($<20$~GeV mass difference) \cite{Dreiner:2012gx,Dreiner:2012sh}, and
more recent studies suggest that the multijet search may actually give stronger constraints on gluino production
as we move to 13~TeV \cite{Chalons:2015vja}.

A region where the 3000~fb$^{-1}$ jets and $E^{\text{miss}}_{\text{T}}$ analysis does provide some 
of the strictest limits is for high mass gluino pair production between 2000 and 2500~GeV (black points). The region
can be more clearly seen in the $m_{\tilde{g}}$ vs $m_{\tilde{t}_1}$ plane of Fig.~\ref{fig:sensitive_analyses} (right)
with the black points at high gluino and stop masses.

The same plot also more clearly illustrates the signal region dependence in the area of parameter
space where stop production dominates. These points can be seen 
for $m_{\tilde{g}} \gsim 2300$~GeV, where the high gluino mass leads to
a decoupling from the LHC phenomenology. Looking at the plot we can clearly see
a difference in the most sensitive analysis for stop masses above and below ${\sim} 1000$~GeV.
Above this mass, the dominant search is the dedicated stop analysis with either 
one lepton (blue points) or 0 leptons (red points) in the final state. Such a result
should be expected as these searches have been precisely tuned for stop masses
in this range.

The green points in Fig.~\ref{fig:sensitive_analyses} (right)
also show that if the
spectrum contains a light sbottom squark, this can be a sensitive production mode. Such a 
search is particularly important for natural SUSY since, as stated before, if the lightest
stop in the spectrum is the SU(2) partner, the corresponding SU(2) sbottom must be close in mass.
We see that for lower masses, the direct search for sbottoms may actually be more sensitive
than those for stops.

\subsection{Distribution of allowed points}

\begin{figure*}[ht] 
\includegraphics[width=0.49\textwidth]{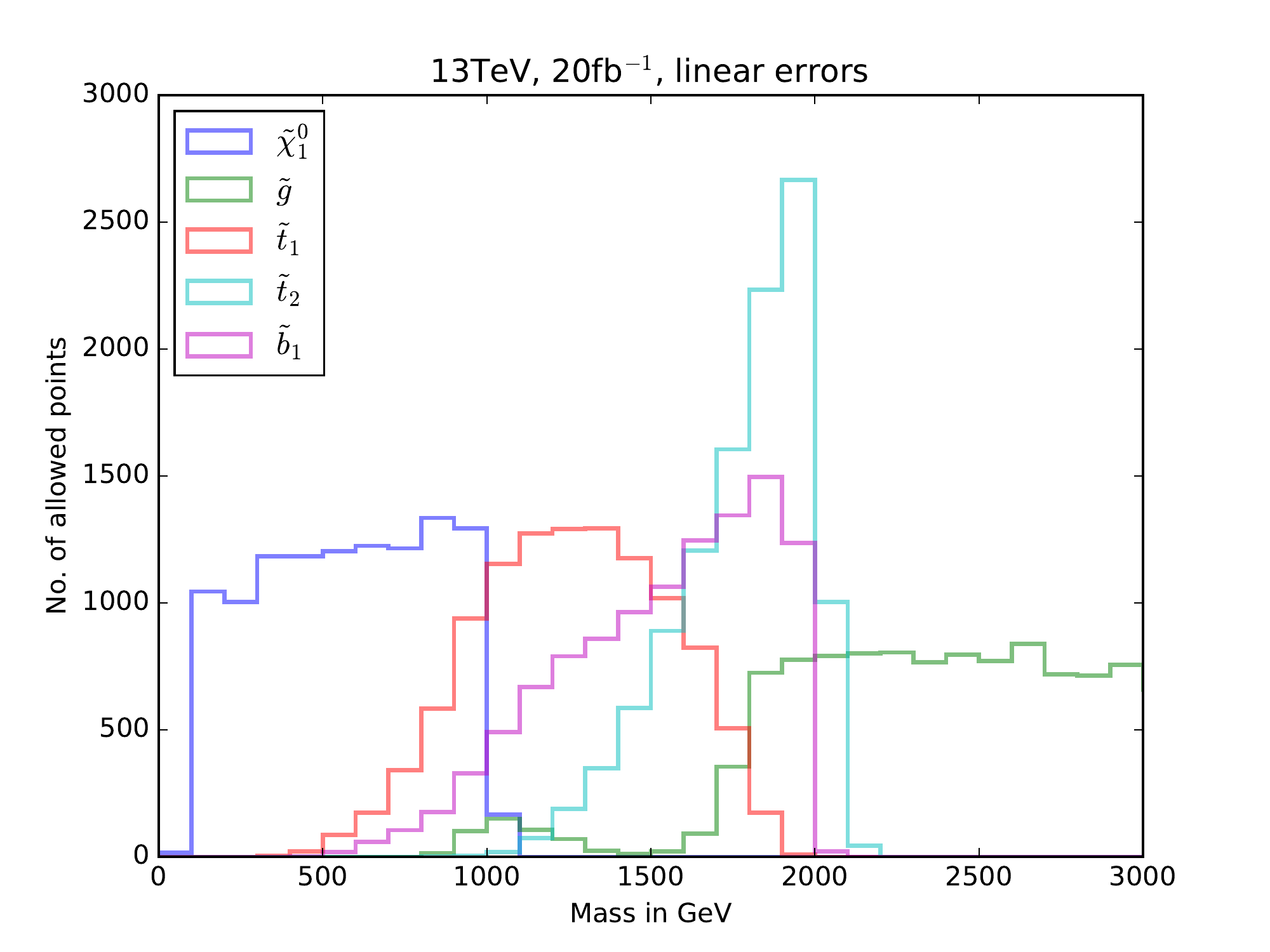} 
\includegraphics[width=0.49\textwidth]{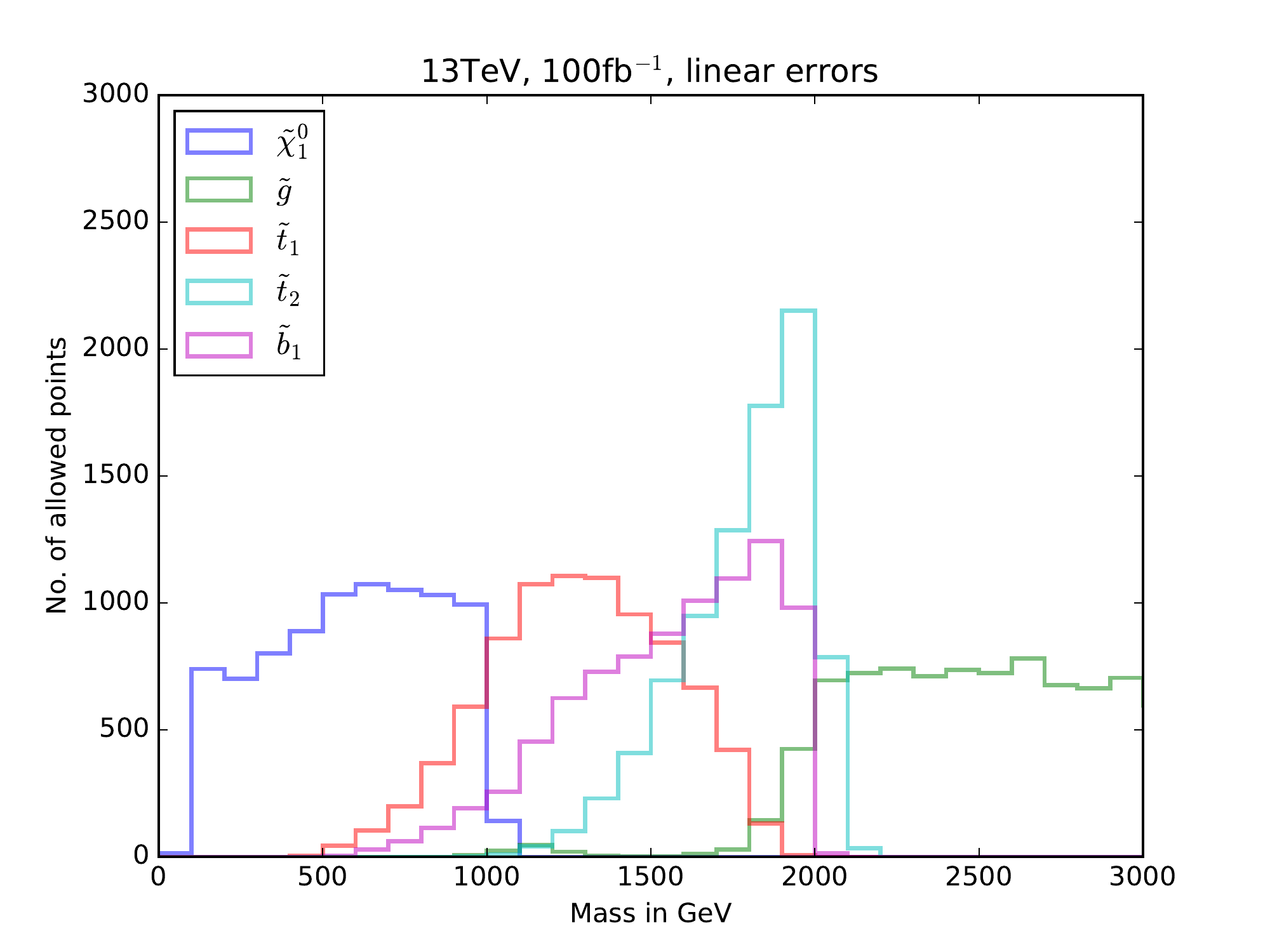}  \\
\includegraphics[width=0.49\textwidth]{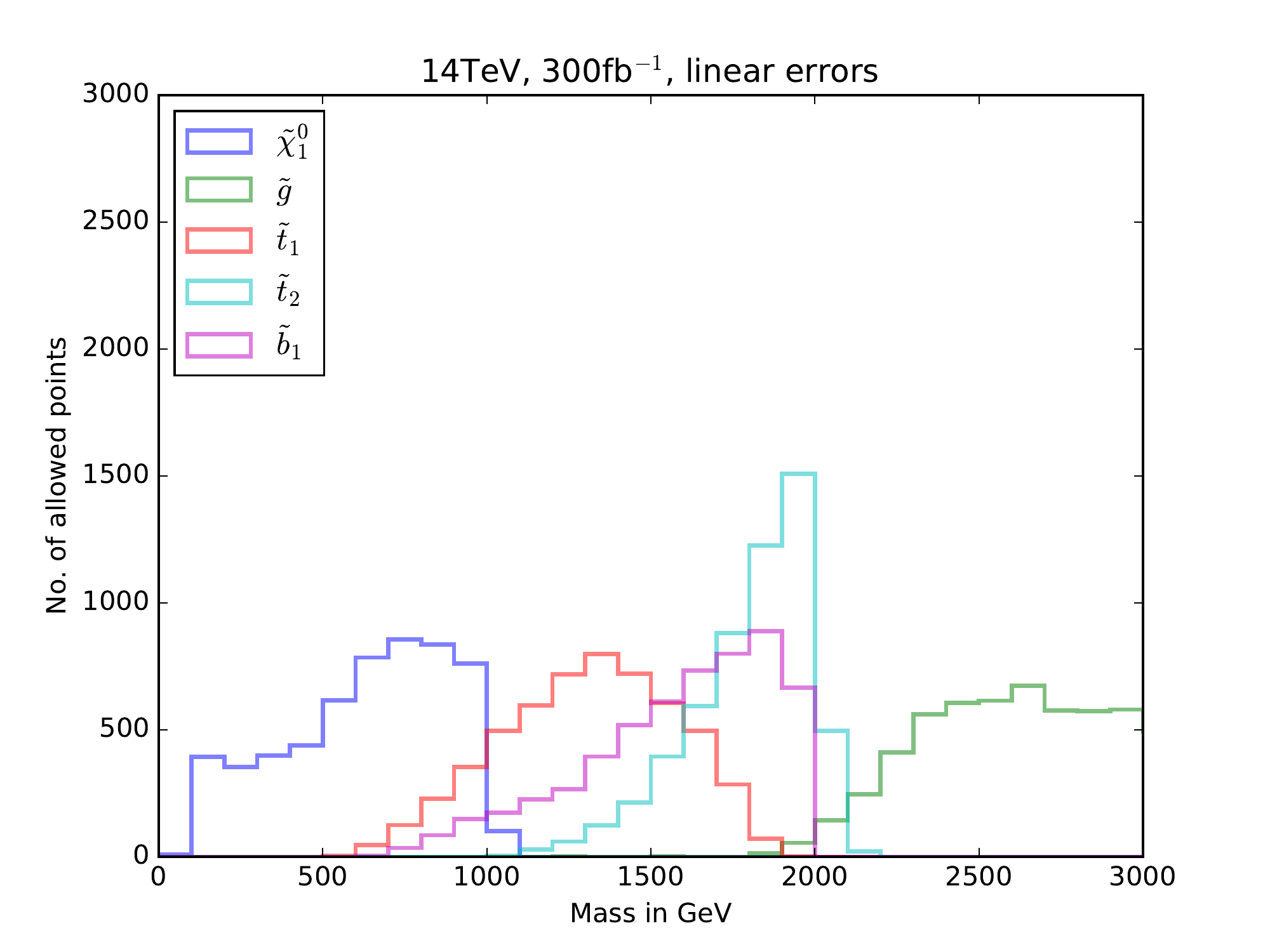}  
\includegraphics[width=0.49\textwidth]{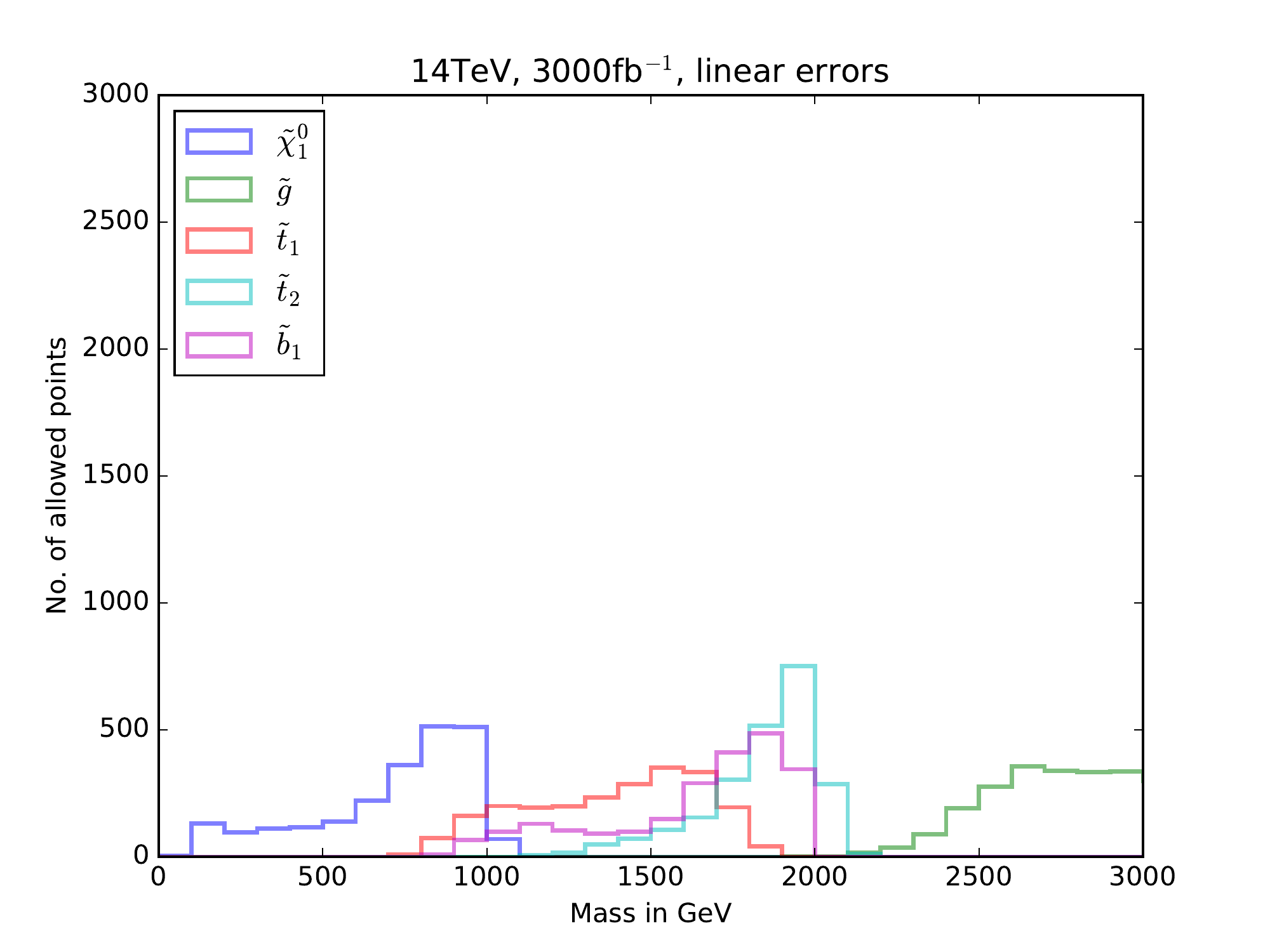} 
\caption{Distribution of particle masses in allowed parameter points after 
13~TeV with 20~fb$^{-1}$ (top left), 13~TeV with 100~fb$^{-1}$ (top right), 
14~TeV with 300~fb$^{-1}$ (bottom left), 
14~TeV with 3000~fb$^{-1}$ (bottom right). }
\label{fig:allowed_hist}  
\end{figure*}

The cumulative effect of the LHC searches at different energies and luminosities
can be most easily seen by plotting distributions of the allowed mass spectra. We show these distributions in Fig.~\ref{fig:allowed_hist} for different 
integrated luminosity and energy stages.

It is instructive to start the analysis by comparing the distributions of all the analysed points, Fig.~\ref{fig:sparticle_mass}, 
to the distribution
after 20~fb$^{-1}$ at $\sqrt{s} = 13$~TeV, the upper left panel of Fig.~\ref{fig:allowed_hist}; we see that roughly 
half of the points become excluded at this stage. The distributions for stops and sbottoms have a similar shape, however,
and this is due to the fact that the majority of the exclusion is driven by the mass of the gluino. 
For higgsinos we observe that the mass distribution is now skewed towards higher masses since compressed
scenarios are more difficult to probe. The 
most dramatic change, however, can be seen for gluinos. The majority of the points 
with $m_{\tilde{g}} \lesssim 1900$~GeV would be excluded at this stage. This confirms the 
fact that gluinos offer the most robust discovery potential at the LHC.

As we continue the analysis for the following LHC stages we observe a rather modest improvement for $\sqrt{s} = 13$~TeV with 100~fb$^{-1}$, the upper right panel of Fig.~\ref{fig:allowed_hist}. There is some reduction in the number of allowed points, of course, which is visible in the $\tilde{t}_2$ distribution. There is also a further shift towards heavier higgsinos. The trend continues for LHC 14~TeV with 300~fb$^{-1}$ (the lower left panel of Fig.~\ref{fig:allowed_hist}), but we also observe an increase in the exclusion limit in the gluino masses, possibly more due to the increased centre-of-mass energy rather than the increased integrated luminosity.

At the final luminosity stage of 3000~fb$^{-1}$ (the lower right panel of Fig.~\ref{fig:allowed_hist}), we see a clear difference in all distributions compared to Fig.~\ref{fig:sparticle_mass}. The allowed points now cumulate around upper limits of the scanning ranges, and the gluino exclusion limit is pushed to ${\sim}2500$~GeV. Not surprisingly, however, there are still allowed points with low higgsino masses that would remain undetected as long as the other superpartners are sufficiently heavy. 

\subsection{Discovery with 3000~fb$^{-1}$ at 14~TeV}

\begin{figure*} 
\includegraphics[width=\textwidth]{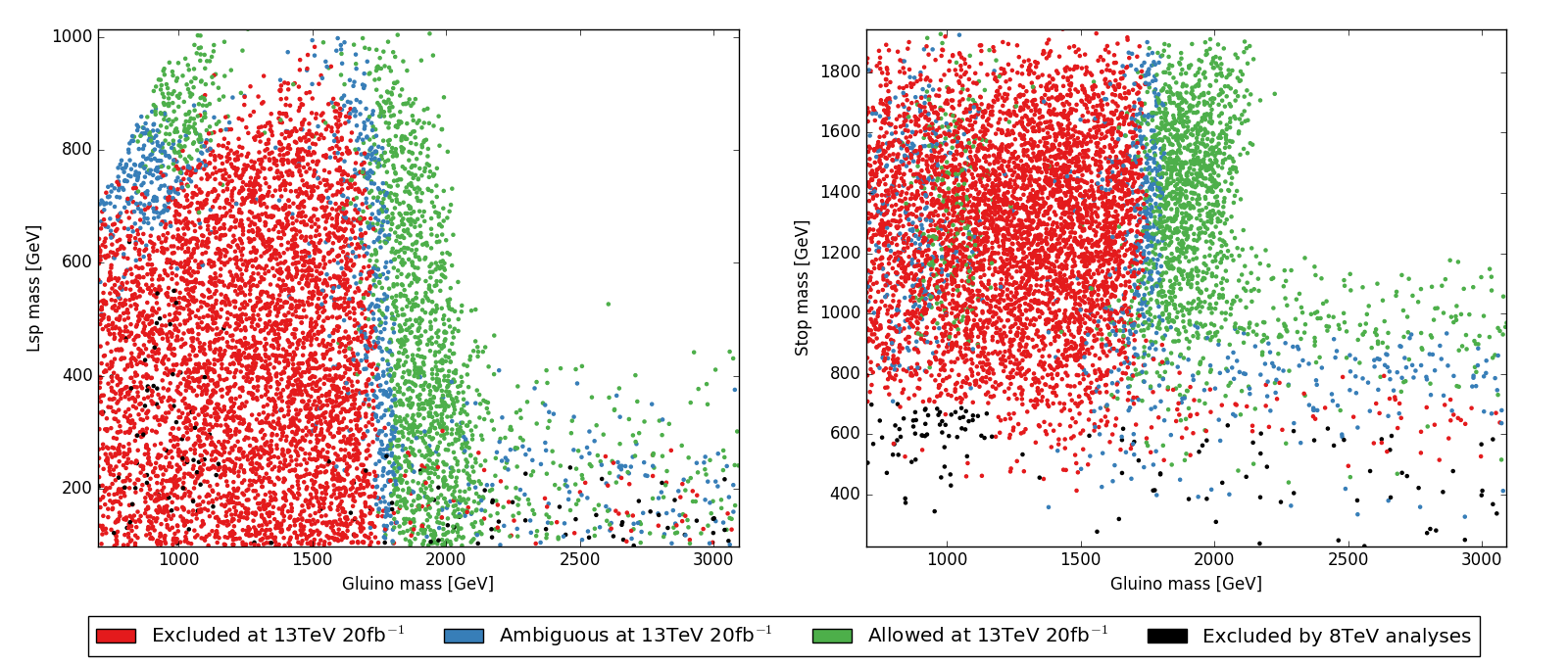} 
\caption{Plots showing the natural SUSY points that can be discovered 
at $\sqrt{s} = \tev{14}$ with $\mathcal{L} = \ifb{3000}$, under the assumption that the current systematic
errors will remain constant. We classify these points as excluded, ambiguous
within Monte Carlo uncertainty and allowed at 13~TeV with 20~fb$^{-1}$, and those
already excluded at 8~TeV.
Left: $m_{\tilde{g}}$ vs $m_{\tilde{\chi}^0_1}$. Right: 
$m_{\tilde{g}}$ vs $m_{\tilde{t}_1}$.}
\label{fig:discover_3000_14}  
\end{figure*} 

Of course, the focus of the LHC is not purely to exclude models of new physics but to hopefully 
actually discover new particles. For this reason, we also make projections for the discovery
potential (5$\sigma$) of the HL LHC at 14~TeV with 3000~fb$^{-1}$. In Fig.~\ref{fig:discover_3000_14} (left)
we plot the points that can be discovered in the $m_{\tilde{g}}$ vs $m_{\tilde{\chi}^0_1}$ plane and colour 
these points according to whether they are predicted to be excluded, ambiguous or allowed with 20~fb$^{-1}$ at 13~TeV.
The plot shows that gluinos up to ${\sim}2000$ GeV can be discovered for light LSPs (the points beyond this are
mainly due to light stops being present in the spectrum). As for the exclusion curves, discovery is much harder
in the compressed region but can be made for some points up to $m_{\tilde{g}}\sim1000$ GeV. 

However, what is far more
striking is that the vast majority of these points would have been excluded with 20~fb$^{-1}$ at 13~TeV. In fact only a 
very small proportion of points (in green) are definitely allowed with 20~fb$^{-1}$ at 13~TeV and these points lie in
two distinct regions. First, there exists a thin strip, $1800 \lesssim m_{\tilde{g}} \lesssim 2100$~GeV where allowed
points can be discovered. Second, in the region of high compression, the much higher statistics allows points to be discovered
that were not excluded with 20~fb$^{-1}$ at 13~TeV. We therefore come to
the conclusion that the coming LHC run will already begin to probe the majority of parameter points that contain 
a gluino within the eventual LHC discovery reach. Essentially, if natural SUSY is to be discovered at the LHC, we
may expect the first signs to appear this year.

A very similar conclusion is reached if we examine the parameter points with light stop masses that can 
be discovered at the HL LHC. In Fig.~\ref{fig:discover_3000_14} (right) we show the
points that can be discovered in the $m_{\tilde{g}}$ vs $m_{\tilde{t}_1}$ plane. Here the majority
of points with $m_{\tilde{g}} < 2000$~GeV are excluded by direct gluino searches, but beyond this, 
we see the parameter points that can be discovered thanks to direct stop production. We see that
stop masses up to ${\sim}1200$~GeV can be discovered in the natural SUSY paradigm. However, we also
see that the majority of the points that can be discovered, will already have been excluded with 20~fb$^{-1}$ at 13~TeV.
Once again, we see only a narrow strip (with some compressed spectra points below), this time 
in the range $800 \lesssim m_{\tilde{g}} \lesssim 1200$~GeV,
which contains points that cannot be probed with 20~fb$^{-1}$ at 13~TeV but can be discovered. Therefore,
for stops we also come to the conclusion that the first signs should appear soon if an unambiguous
discovery can be made at the LHC.

\subsection{Discovery with 3000~fb$^{-1}$ at 14~TeV: Reduced~errors}

\begin{figure*} 
\includegraphics[width=\textwidth]{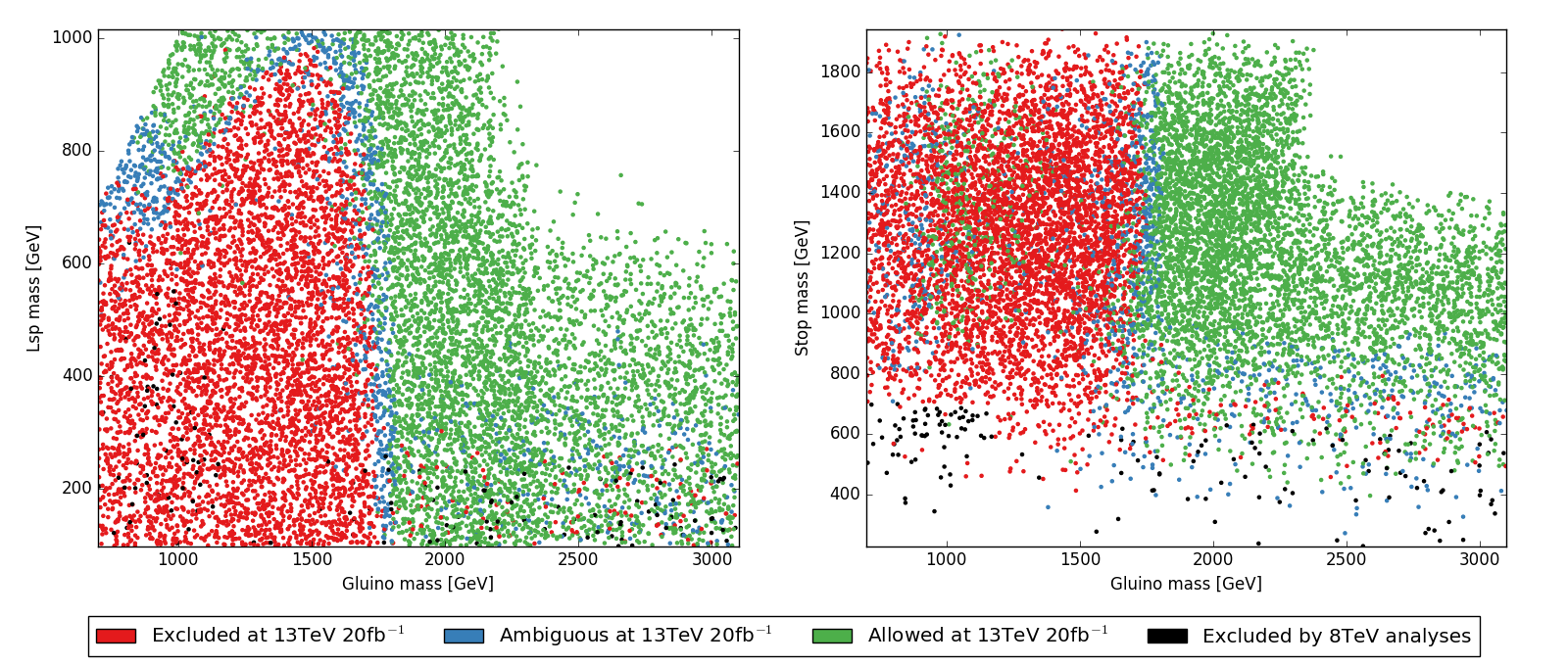} 
\caption{Plots showing the natural SUSY points that can be discovered 
at $\sqrt{s} = \tev{14}$ with $\mathcal{L} = \ifb{3000}$, under the assumption that the systematic
errors will reduce in proportion to the collected luminosity. We classify these points as excluded, ambiguous
within MC uncertainty and allowed at 13~TeV with 20~fb$^{-1}$, and those
already excluded at 8~TeV.
Left: $m_{\tilde{g}}$ vs $m_{\tilde{\chi}^0_1}$. 
Right: $m_{\tilde{g}}$ vs $m_{\tilde{t}_1}$.}
\label{fig:discover_3000_14_sqrtErr}  
\end{figure*}

The previous section makes the assumption that the systematic uncertainties 
on the background and signal remain at their current values. However, we 
also investigate how the discovery
potential of the LHC is improved if the uncertainties can be reduced. We therefore
again reduce the systematic uncertainties according to the increase in the collected
statistics in the same way as Sec.~\ref{sec:exclu_3000_14_sqrtErr}.

If we now compare the previously discussed discovery potential in Fig.~\ref{fig:discover_3000_14} 
with the reduced uncertainty case shown in Fig.~\ref{fig:discover_3000_14_sqrtErr}, we see a
significant increase in the SUSY masses that the LHC can discover. Specifically, we see that 
in a natural SUSY setup we can expect to discover gluino masses up to $m_{\tilde{g}} \sim 2400$~GeV. Perhaps
more importantly, however, is the fact that Fig.~\ref{fig:discover_3000_14_sqrtErr} (left) shows that the coverage 
for lighter gluinos ($m_{\tilde{g}} \lesssim 1500$~GeV) but heavier LSPs ($m_{\tilde{\chi}^0_1} \gsim 800$~GeV)
is now far more comprehensive. Nevertheless, for the majority of these points, the first signs should already
be seen at 13~TeV with 20~fb$^{-1}$.

A similar conclusion can also be reached when the discovery is facilitated by light top squarks in the 
spectrum. Examining Fig.~\ref{fig:discover_3000_14_sqrtErr} (right), we see that stops up to 
$m_{\tilde{t}_1} \sim 1400$~GeV can now be discovered, and this is to be compared to ${\sim} 1200$~GeV if the systematic
errors are not reduced. As for the gluino case, however, the increase in the discovery potential is perhaps
undersold by only studying the mass reach. More important is the fact that without reduced uncertainties, only a very small
region of parameter space can be discovered with 3000~fb$^{-1}$ at 14~TeV that is not already ruled out with 20~fb$^{-1}$ 
at 13~TeV. If the uncertainties can be reduced, however, this picture changes significantly, and far more of 
the parameter space is left open for discovery.

\section{Summary and Conclusion}
\label{sec:summary} 
In this paper we have examined the prospects of probing and even discovering 
so-called ``natural'' SUSY as the LHC progresses in both collected luminosity and 
energy. To map the progress we considered four different LHC scenarios: 
20~fb$^{-1}$ at 13~TeV, 100~fb$^{-1}$ at 13~TeV, 300~fb$^{-1}$ at 14~TeV, and
3000~fb$^{-1}$ at 14~TeV, and first explored the gluino and lightest stop
masses that can be excluded at each setup.

For gluino masses, the maximum exclusion ranges from $\sim$1.7~TeV with 20~fb$^{-1}$ at 13~TeV to
$\sim$2.5~TeV with 3000~fb$^{-1}$ at 14~TeV. In contrast, the lower production cross sections
for stops result in correspondingly weaker bounds of $\sim$800~GeV with 20~fb$^{-1}$ at 13~TeV to
$\sim$1.5~TeV with 3000~fb$^{-1}$ at 14~TeV. We should also mention that these bounds heavily
depend on the mass of the LSP in the spectrum. In particular, as the LSP mass rises, the spectra
become compressed, and we see a reduction in the LHC reach. For example, with 3000~fb$^{-1}$ at 14~TeV
we can set no bound on the stop mass if $\tilde{\chi}^0_1\gsim600$~GeV, but many points also survive 
with both lower stop and LSP masses.

The above conclusions assumed that the systematic uncertainties remain at approximately their current levels. In
order to see how the LHC reach depends on this assumption, we also investigated the effect of reducing 
the systematic uncertainty according to the collected luminosity (i.e.\ the uncertainty scales as $1/\sqrt{\mathcal{L}}$). 
Reducing the uncertainty in such a way only has a small effect on the LHC reach for gluinos, and the mass
limits rise by ${\sim}100$~GeV to $m_{\tilde{g}}\gsim2600$. The small difference is due to the fact that for such
high masses, gluino production is almost at the kinematic limit, and thus reduced systematic errors can only
marginally improve the bound. In contrast, the limits on stop production are markedly improved. In particular,
for low mass LSP stops can be reliably excluded for $m_{\tilde{t}_1} < 1400$~GeV, whereas if the uncertainties remain
constant, many points with $m_{\tilde{t}_1}\sim1000$~GeV are ambiguous. A similar conclusion can be seen in terms of the
LSP mass, where with the reduced errors, very few points with $m_{\tilde{\chi}^0_1}<700$~GeV remain (assuming $m_{\tilde{t}_1}<1400$~GeV).
This can be compared to the case without reduced errors, where it is very hard to make strong statements in
the $m_{\tilde{t}_1}$ vs $m_{\tilde{\chi}^0_1}$ plane.

Our most important results, however, are regarding the expected discovery reach of the LHC at 
high luminosity (3000~fb$^{-1}$ at 14~TeV). We come to the rather sobering conclusion that assuming 
the systematic errors remain constant, there are relatively few parameter points that can be discovered that
are not already excluded with 20~fb$^{-1}$ at 13~TeV. Essentially, if natural SUSY is to be discovered at the
LHC, the first hints probably need to start appearing this summer. The situation changes slightly if the systematic
errors can be reduced as more data are collected at the LHC. If we again take the scenario that
the systematic uncertainty scales as $1/\sqrt{\mathcal{L}}$, the prospects of a natural SUSY discovery
at high luminosity become more optimistic. For gluinos we find a mass band $1800\lsim m_{\tilde{g}}\lsim2400$,
where spectra that are not excluded with 20~fb$^{-1}$ at 13~TeV can still eventually be discovered. A similar
mass band $900\lsim m_{\tilde{t}_1}\lsim1400$ exists for stops, and this shows the importance of reducing 
the systematic uncertainties to fully exploit the discovery potential of the high luminosity LHC.

 
\begin{acknowledgments} 
The authors thank Sascha Caron for useful discussions. The work of JSK has been partially supported by the MINECO, Spain, under Contract No.\ FPA2013-44773-P; Consolider-Ingenio CPAN CSD2007-00042 and  the Spanish MINECO Centro de Excelencia Severo Ochoa Program under Grant No.\ SEV-2012-0249. JSK would like to thank the IBS CPTU in Daejeon for support and hospitality while part of this manuscript was prepared. KR is supported by the National Science Centre (Poland) under Grant No.\ 2015/19/D/ST2/03136.
\end{acknowledgments} 

\appendix

\bibliographystyle{utphys}
\bibliography{natural2}

\providecommand{\href}[2]{#2}\begingroup\raggedright\begin{thebibliography}{10}

\bibitem{Aad:2015baa}
{\bf ATLAS} Collaboration, G.~Aad {\em et al.}, ``{Summary of the ATLAS
  experiment’s sensitivity to supersymmetry after LHC Run 1 — interpreted
  in the phenomenological MSSM},''
  \href{http://dx.doi.org/10.1007/JHEP10(2015)134}{{\em JHEP} {\bf 10} (2015)
  134},
\href{http://arxiv.org/abs/1508.06608}{{\tt arXiv:1508.06608 [hep-ex]}}.

\bibitem{Khachatryan:2016nvf}
{\bf CMS} Collaboration, V.~Khachatryan {\em et al.}, ``{Phenomenological MSSM
  interpretation of CMS searches in $pp$ collisions at $\sqrt{s} = 7$ and 8
  TeV},'' \href{http://dx.doi.org/10.1007/JHEP10(2016)129}{{\em JHEP} {\bf 10}
  (2016)  129},
\href{http://arxiv.org/abs/1606.03577}{{\tt arXiv:1606.03577 [hep-ex]}}.

\bibitem{Aad:2015pfx}
{\bf ATLAS} Collaboration, G.~Aad {\em et al.}, ``{ATLAS Run 1 searches for
  direct pair production of third-generation squarks at the Large Hadron
  Collider},'' \href{http://dx.doi.org/10.1140/epjc/s10052-015-3726-9,
  10.1140/epjc/s10052-016-3935-x}{{\em Eur. Phys. J.} {\bf C75} (2015) no.~10,
  510}, \href{http://arxiv.org/abs/1506.08616}{{\tt arXiv:1506.08616
  [hep-ex]}}.
[Erratum: Eur. Phys. J.C76,no.3,153(2016)].

\bibitem{Aad:2015iea}
{\bf ATLAS} Collaboration, G.~Aad {\em et al.}, ``{Summary of the searches for
  squarks and gluinos using $ \sqrt{s}=8 $ TeV pp collisions with the ATLAS
  experiment at the LHC},''
  \href{http://dx.doi.org/10.1007/JHEP10(2015)054}{{\em JHEP} {\bf 10} (2015)
  054},
\href{http://arxiv.org/abs/1507.05525}{{\tt arXiv:1507.05525 [hep-ex]}}.

\bibitem{Chatrchyan:2013iqa}
{\bf CMS} Collaboration, S.~Chatrchyan {\em et al.}, ``{Search for
  supersymmetry in pp collisions at $\sqrt{s}=8$ TeV in events with a single
  lepton, large jet multiplicity, and multiple $b$ jets},''
  \href{http://dx.doi.org/10.1016/j.physletb.2014.04.023}{{\em Phys. Lett.}
  {\bf B733} (2014)  328--353},
\href{http://arxiv.org/abs/1311.4937}{{\tt arXiv:1311.4937 [hep-ex]}}.

\bibitem{Khachatryan:2016oia}
{\bf CMS} Collaboration, V.~Khachatryan {\em et al.}, ``{Search for direct pair
  production of supersymmetric top quarks decaying to all-hadronic final states
  in $pp$ collisions at $\sqrt{s} = 8\;\text {TeV} $},''
  \href{http://dx.doi.org/10.1140/epjc/s10052-016-4292-5}{{\em Eur. Phys. J.}
  {\bf C76} (2016) no.~8, 460},
\href{http://arxiv.org/abs/1603.00765}{{\tt arXiv:1603.00765 [hep-ex]}}.

\bibitem{Chamseddine:1982jx}
A.~H. Chamseddine, R.~L. Arnowitt, and P.~Nath, ``{Locally Supersymmetric Grand
  Unification},''
\href{http://dx.doi.org/10.1103/PhysRevLett.49.970}{{\em Phys. Rev. Lett.} {\bf
  49} (1982)  970}.

\bibitem{Feng:1999mn}
J.~L. Feng, K.~T. Matchev, and T.~Moroi, ``{Multi - TeV scalars are natural in
  minimal supergravity},''
  \href{http://dx.doi.org/10.1103/PhysRevLett.84.2322}{{\em Phys. Rev. Lett.}
  {\bf 84} (2000)  2322--2325},
\href{http://arxiv.org/abs/hep-ph/9908309}{{\tt arXiv:hep-ph/9908309
  [hep-ph]}}.

\bibitem{Kitano:2006gv}
R.~Kitano and Y.~Nomura, ``{Supersymmetry, naturalness, and signatures at the
  LHC},'' \href{http://dx.doi.org/10.1103/PhysRevD.73.095004}{{\em Phys. Rev.}
  {\bf D73} (2006)  095004},
\href{http://arxiv.org/abs/hep-ph/0602096}{{\tt arXiv:hep-ph/0602096
  [hep-ph]}}.

\bibitem{Baer:2012uy}
H.~Baer, V.~Barger, P.~Huang, and X.~Tata, ``{Natural Supersymmetry: LHC, dark
  matter and ILC searches},''
  \href{http://dx.doi.org/10.1007/JHEP05(2012)109}{{\em JHEP} {\bf 05} (2012)
  109},
\href{http://arxiv.org/abs/1203.5539}{{\tt arXiv:1203.5539 [hep-ph]}}.

\bibitem{Baer:2011ec}
H.~Baer, V.~Barger, and P.~Huang, ``{Hidden SUSY at the LHC: the light
  higgsino-world scenario and the role of a lepton collider},''
  \href{http://dx.doi.org/10.1007/JHEP11(2011)031}{{\em JHEP} {\bf 11} (2011)
  031},
\href{http://arxiv.org/abs/1107.5581}{{\tt arXiv:1107.5581 [hep-ph]}}.

\bibitem{Papucci:2011wy}
M.~Papucci, J.~T. Ruderman, and A.~Weiler, ``{Natural SUSY Endures},''
  \href{http://dx.doi.org/10.1007/JHEP09(2012)035}{{\em JHEP} {\bf 1209} (2012)
   035},
\href{http://arxiv.org/abs/1110.6926}{{\tt arXiv:1110.6926 [hep-ph]}}.

\bibitem{Barbieri198863}
R.~Barbieri and G.~Giudice, ``Upper bounds on supersymmetric particle masses,''
  \href{http://dx.doi.org/http://dx.doi.org/10.1016/0550-3213(88)90171-X}{{\em
  Nuclear Physics B} {\bf 306} (1988) no.~1, 63 -- 76}.
  \url{http://www.sciencedirect.com/science/article/pii/055032138890171X}.

\bibitem{Ellis:1986yg}
J.~R. Ellis, K.~Enqvist, D.~V. Nanopoulos, and F.~Zwirner, ``{Observables in
  Low-Energy Superstring Models},''
\href{http://dx.doi.org/10.1142/S0217732386000105}{{\em Mod. Phys. Lett.} {\bf
  A01} (1986)  57}.

\bibitem{Casas:2014eca}
J.~A. Casas, J.~M. Moreno, S.~Robles, K.~Rolbiecki, and B.~Zaldívar, ``{What
  is a Natural SUSY scenario?},''
  \href{http://dx.doi.org/10.1007/JHEP06(2015)070}{{\em JHEP} {\bf 06} (2015)
  070},
\href{http://arxiv.org/abs/1407.6966}{{\tt arXiv:1407.6966 [hep-ph]}}.

\bibitem{Han:2013kga}
C.~Han, K.-i. Hikasa, L.~Wu, J.~M. Yang, and Y.~Zhang, ``{Current experimental
  bounds on stop mass in natural SUSY},''
  \href{http://dx.doi.org/10.1007/JHEP10(2013)216}{{\em JHEP} {\bf 10} (2013)
  216},
\href{http://arxiv.org/abs/1308.5307}{{\tt arXiv:1308.5307 [hep-ph]}}.

\bibitem{Kowalska:2013ica}
K.~Kowalska and E.~M. Sessolo, ``{Natural MSSM after the LHC 8 TeV run},''
  \href{http://dx.doi.org/10.1103/PhysRevD.88.075001}{{\em Phys. Rev.} {\bf
  D88} (2013) no.~7, 075001},
\href{http://arxiv.org/abs/1307.5790}{{\tt arXiv:1307.5790 [hep-ph]}}.

\bibitem{Buchmueller:2013exa}
O.~Buchmueller and J.~Marrouche, ``{Universal mass limits on gluino and
  third-generation squarks in the context of Natural-like SUSY spectra},''
  \href{http://dx.doi.org/10.1142/S0217751X14500328}{{\em Int. J. Mod. Phys.}
  {\bf A29} (2014) no.~06, 1450032},
\href{http://arxiv.org/abs/1304.2185}{{\tt arXiv:1304.2185 [hep-ph]}}.

\bibitem{Belanger:2015vwa}
G.~Belanger, D.~Ghosh, R.~Godbole, and S.~Kulkarni, ``{Light stop in the MSSM
  after LHC Run 1},'' \href{http://dx.doi.org/10.1007/JHEP09(2015)214}{{\em
  JHEP} {\bf 09} (2015)  214},
\href{http://arxiv.org/abs/1506.00665}{{\tt arXiv:1506.00665 [hep-ph]}}.

\bibitem{Kobakhidze:2015scd}
A.~Kobakhidze, N.~Liu, L.~Wu, J.~M. Yang, and M.~Zhang, ``{Closing up a light
  stop window in natural SUSY at LHC},''
  \href{http://dx.doi.org/10.1016/j.physletb.2016.02.003}{{\em Phys. Lett.}
  {\bf B755} (2016)  76--81},
\href{http://arxiv.org/abs/1511.02371}{{\tt arXiv:1511.02371 [hep-ph]}}.

\bibitem{Kim:2015dpa}
J.~S. Kim, D.~Schmeier, and J.~Tattersall, ``{Role of the ‘N’ in the
  natural NMSSM for the LHC},''
  \href{http://dx.doi.org/10.1103/PhysRevD.93.055018}{{\em Phys. Rev.} {\bf
  D93} (2016) no.~5, 055018},
\href{http://arxiv.org/abs/1510.04871}{{\tt arXiv:1510.04871 [hep-ph]}}.

\bibitem{Drees:2015aeo}
M.~Drees and J.~S. Kim, ``{Minimal natural supersymmetry after the LHC8},''
  \href{http://dx.doi.org/10.1103/PhysRevD.93.095005}{{\em Phys. Rev.} {\bf
  D93} (2016) no.~9, 095005},
\href{http://arxiv.org/abs/1511.04461}{{\tt arXiv:1511.04461 [hep-ph]}}.

\bibitem{Aad:2015zhl}
{\bf ATLAS, CMS} Collaboration, G.~Aad {\em et al.}, ``{Combined Measurement of
  the Higgs Boson Mass in $pp$ Collisions at $\sqrt{s}=7$ and 8 TeV with the
  ATLAS and CMS Experiments},''
  \href{http://dx.doi.org/10.1103/PhysRevLett.114.191803}{{\em Phys. Rev.
  Lett.} {\bf 114} (2015)  191803},
\href{http://arxiv.org/abs/1503.07589}{{\tt arXiv:1503.07589 [hep-ex]}}.

\bibitem{Bechtle:2012jw}
P.~Bechtle, S.~Heinemeyer, O.~Stal, T.~Stefaniak, G.~Weiglein, and L.~Zeune,
  ``{MSSM Interpretations of the LHC Discovery: Light or Heavy Higgs?},''
  \href{http://dx.doi.org/10.1140/epjc/s10052-013-2354-5}{{\em Eur. Phys. J.}
  {\bf C73} (2013) no.~4, 2354},
\href{http://arxiv.org/abs/1211.1955}{{\tt arXiv:1211.1955 [hep-ph]}}.

\bibitem{lhc-prospects}
J.~Wenniger, ``{LHC operation in 2015 and prospects for the future}.'' {Moriond
  - La Thuile}, 2016.
\newblock
  \url{https://indico.in2p3.fr/event/12279/session/7/contribution/118/material/slides/0.pdf}.

\bibitem{lhc-prospects2}
``{LHC commissioning schedule}.''
\newblock
  \url{http://lhc-commissioning.web.cern.ch/lhc-commissioning/schedule/LHC-schedule-update.pdf}.
  [Online; accessed 20-May-2016].

\bibitem{Ibanez:1991pr}
L.~E. Ibanez and G.~G. Ross, ``{Discrete gauge symmetries and the origin of
  baryon and lepton number conservation in supersymmetric versions of the
  standard model},''
\href{http://dx.doi.org/10.1016/0550-3213(92)90195-H}{{\em Nucl. Phys.} {\bf
  B368} (1992)  3--37}.

\bibitem{Drees:2013wra}
M.~Drees, H.~Dreiner, D.~Schmeier, J.~Tattersall, and J.~S. Kim, ``{CheckMATE:
  Confronting your Favourite New Physics Model with LHC Data},''
  \href{http://dx.doi.org/10.1016/j.cpc.2014.10.018}{{\em Comput. Phys.
  Commun.} {\bf 187} (2015)  227--265},
\href{http://arxiv.org/abs/1312.2591}{{\tt arXiv:1312.2591 [hep-ph]}}.

\bibitem{deFavereau:2013fsa}
{\bf DELPHES 3} Collaboration, J.~de~Favereau {\em et al.}, ``{DELPHES 3, A
  modular framework for fast simulation of a generic collider experiment},''
  \href{http://dx.doi.org/10.1007/JHEP02(2014)057}{{\em JHEP} {\bf 1402} (2014)
   057},
\href{http://arxiv.org/abs/1307.6346}{{\tt arXiv:1307.6346 [hep-ex]}}.

\bibitem{LEPconst}
``{The LEP SUSY Working Group and the ALEPH, DELPHI, L3 and OPAL experiments,
  note LEPSUSYWG/01-03.1}.''
\newblock \url{http://lepsusy.web.cern.ch/lepsusy}.

\bibitem{Ross:2016pml}
G.~G. Ross, K.~Schmidt-Hoberg, and F.~Staub, ``{On the MSSM Higgsino mass and
  fine tuning},'' \href{http://dx.doi.org/10.1016/j.physletb.2016.05.053}{{\em
  Phys. Lett.} {\bf B759} (2016)  110--114},
\href{http://arxiv.org/abs/1603.09347}{{\tt arXiv:1603.09347 [hep-ph]}}.

\bibitem{Kim:2015hda}
J.~S. Kim, O.~Lebedev, and D.~Schmeier, ``{Higgsophilic gauge bosons and
  monojets at the LHC},'' \href{http://dx.doi.org/10.1007/JHEP11(2015)128}{{\em
  JHEP} {\bf 11} (2015)  128},
\href{http://arxiv.org/abs/1507.08673}{{\tt arXiv:1507.08673 [hep-ph]}}.

\bibitem{PhysRevD.89.055007}
H.~Baer, A.~Mustafayev, and X.~Tata,
  \href{http://dx.doi.org/10.1103/PhysRevD.89.055007}{``Monojets and
  monophotons from light higgsino pair production at lhc14,''{\em Phys. Rev. D}
  {\bf 89} (Mar, 2014)  055007}.
  \url{http://link.aps.org/doi/10.1103/PhysRevD.89.055007}.

\bibitem{Rolbiecki:2009hk}
K.~Rolbiecki, J.~Tattersall, and G.~Moortgat-Pick, ``{Towards Measuring the
  Stop Mixing Angle at the LHC},''
  \href{http://dx.doi.org/10.1140/epjc/s10052-010-1517-x}{{\em Eur. Phys. J.}
  {\bf C71} (2011)  1517},
\href{http://arxiv.org/abs/0909.3196}{{\tt arXiv:0909.3196 [hep-ph]}}.

\bibitem{Hikasa:1987db}
K.-i. Hikasa and M.~Kobayashi, ``{Light Scalar Top at $e^+ e^-$ Colliders},''
\href{http://dx.doi.org/10.1103/PhysRevD.36.724}{{\em Phys. Rev.} {\bf D36}
  (1987)  724}.

\bibitem{Baer:1990sc}
H.~Baer, X.~Tata, and J.~Woodside, ``{Phenomenology of Gluino Decays via Loops
  and Top Quark Yukawa Coupling},''
\href{http://dx.doi.org/10.1103/PhysRevD.42.1568}{{\em Phys. Rev.} {\bf D42}
  (1990)  1568--1576}.

\bibitem{Porod:2011nf}
W.~Porod and F.~Staub, ``{SPheno 3.1: Extensions including flavour, CP-phases
  and models beyond the MSSM},''
  \href{http://dx.doi.org/10.1016/j.cpc.2012.05.021}{{\em Comput. Phys.
  Commun.} {\bf 183} (2012)  2458--2469},
\href{http://arxiv.org/abs/1104.1573}{{\tt arXiv:1104.1573 [hep-ph]}}.

\bibitem{Sjostrand:2014zea}
T.~Sjöstrand, S.~Ask, J.~R. Christiansen, R.~Corke, N.~Desai, P.~Ilten,
  S.~Mrenna, S.~Prestel, C.~O. Rasmussen, and P.~Z. Skands, ``{An Introduction
  to PYTHIA 8.2},'' \href{http://dx.doi.org/10.1016/j.cpc.2015.01.024}{{\em
  Comput. Phys. Commun.} {\bf 191} (2015)  159--177},
\href{http://arxiv.org/abs/1410.3012}{{\tt arXiv:1410.3012 [hep-ph]}}.

\bibitem{Desai:2011su}
N.~Desai and P.~Z. Skands, ``{Supersymmetry and Generic BSM Models in PYTHIA
  8},'' \href{http://dx.doi.org/10.1140/epjc/s10052-012-2238-0}{{\em Eur. Phys.
  J.} {\bf C72} (2012)  2238},
\href{http://arxiv.org/abs/1109.5852}{{\tt arXiv:1109.5852 [hep-ph]}}.

\bibitem{Nadolsky:2008zw}
P.~M. Nadolsky, H.-L. Lai, Q.-H. Cao, J.~Huston, J.~Pumplin, D.~Stump, W.-K.
  Tung, and C.~P. Yuan, ``{Implications of CTEQ global analysis for collider
  observables},'' \href{http://dx.doi.org/10.1103/PhysRevD.78.013004}{{\em
  Phys. Rev.} {\bf D78} (2008)  013004},
\href{http://arxiv.org/abs/0802.0007}{{\tt arXiv:0802.0007 [hep-ph]}}.

\bibitem{Beenakker:1996ch}
W.~Beenakker, R.~Hopker, M.~Spira, and P.~Zerwas, ``{Squark and gluino
  production at hadron colliders},''
  \href{http://dx.doi.org/10.1016/S0550-3213(97)80027-2}{{\em Nucl.Phys.} {\bf
  B492} (1997)  51--103},
\href{http://arxiv.org/abs/hep-ph/9610490}{{\tt arXiv:hep-ph/9610490
  [hep-ph]}}.

\bibitem{Beenakker:1997ut}
W.~Beenakker, M.~Kramer, T.~Plehn, M.~Spira, and P.~Zerwas, ``{Stop production
  at hadron colliders},''
  \href{http://dx.doi.org/10.1016/S0550-3213(98)00014-5}{{\em Nucl.Phys.} {\bf
  B515} (1998)  3--14},
\href{http://arxiv.org/abs/hep-ph/9710451}{{\tt arXiv:hep-ph/9710451
  [hep-ph]}}.

\bibitem{Kulesza:2008jb}
A.~Kulesza and L.~Motyka, ``{Threshold resummation for squark-antisquark and
  gluino-pair production at the LHC},''
  \href{http://dx.doi.org/10.1103/PhysRevLett.102.111802}{{\em Phys.Rev.Lett.}
  {\bf 102} (2009)  111802},
\href{http://arxiv.org/abs/0807.2405}{{\tt arXiv:0807.2405 [hep-ph]}}.

\bibitem{Kulesza:2009kq}
A.~Kulesza and L.~Motyka, ``{Soft gluon resummation for the production of
  gluino-gluino and squark-antisquark pairs at the LHC},''
  \href{http://dx.doi.org/10.1103/PhysRevD.80.095004}{{\em Phys.Rev.} {\bf D80}
  (2009)  095004},
\href{http://arxiv.org/abs/0905.4749}{{\tt arXiv:0905.4749 [hep-ph]}}.

\bibitem{Beenakker:2010nq}
W.~Beenakker, S.~Brensing, M.~Kramer, A.~Kulesza, E.~Laenen, {\em et al.},
  ``{Supersymmetric top and bottom squark production at hadron colliders},''
  \href{http://dx.doi.org/10.1007/JHEP08(2010)098}{{\em JHEP} {\bf 1008} (2010)
   098},
\href{http://arxiv.org/abs/1006.4771}{{\tt arXiv:1006.4771 [hep-ph]}}.

\bibitem{Beenakker:2011fu}
W.~Beenakker, S.~Brensing, M.~Kramer, A.~Kulesza, E.~Laenen, {\em et al.},
  ``{Squark and Gluino Hadroproduction},''
  \href{http://dx.doi.org/10.1142/S0217751X11053560}{{\em Int.J.Mod.Phys.} {\bf
  A26} (2011)  2637--2664},
\href{http://arxiv.org/abs/1105.1110}{{\tt arXiv:1105.1110 [hep-ph]}}.

\bibitem{Kim:2015wza}
J.~S. Kim, D.~Schmeier, J.~Tattersall, and K.~Rolbiecki, ``{A framework to
  create customised LHC analyses within CheckMATE},''
  \href{http://dx.doi.org/10.1016/j.cpc.2015.06.002}{{\em Comput. Phys.
  Commun.} {\bf 196} (2015)  535--562},
\href{http://arxiv.org/abs/1503.01123}{{\tt arXiv:1503.01123 [hep-ph]}}.

\bibitem{webpage}
\url{{http://checkmate.hepforge.org/}}.

\bibitem{Cacciari:2005hq}
M.~Cacciari and G.~P. Salam, ``{Dispelling the $N^{3}$ myth for the $k_t$
  jet-finder},'' \href{http://dx.doi.org/10.1016/j.physletb.2006.08.037}{{\em
  Phys.Lett.} {\bf B641} (2006)  57--61},
\href{http://arxiv.org/abs/hep-ph/0512210}{{\tt arXiv:hep-ph/0512210
  [hep-ph]}}.

\bibitem{Cacciari:2008gp}
M.~Cacciari, G.~P. Salam, and G.~Soyez, ``{The Anti-$k_t$ jet clustering
  algorithm},'' \href{http://dx.doi.org/10.1088/1126-6708/2008/04/063}{{\em
  JHEP} {\bf 0804} (2008)  063},
\href{http://arxiv.org/abs/0802.1189}{{\tt arXiv:0802.1189 [hep-ph]}}.

\bibitem{Cacciari:2011ma}
M.~Cacciari, G.~P. Salam, and G.~Soyez, ``{FastJet User Manual},''
  \href{http://dx.doi.org/10.1140/epjc/s10052-012-1896-2}{{\em Eur.Phys.J.}
  {\bf C72} (2012)  1896},
\href{http://arxiv.org/abs/1111.6097}{{\tt arXiv:1111.6097 [hep-ph]}}.

\bibitem{Drees:1990dx}
M.~Drees and K.~Hagiwara, ``{Supersymmetric Contribution to the Electroweak
  $\rho$ Parameter},''
\href{http://dx.doi.org/10.1103/PhysRevD.42.1709}{{\em Phys. Rev.} {\bf D42}
  (1990)  1709--1725}.

\bibitem{Amhis:2012bh}
{\bf Heavy Flavor Averaging Group} Collaboration, Y.~Amhis {\em et al.},
  ``{Averages of B-Hadron, C-Hadron, and tau-lepton properties as of early
  2012},''
\href{http://arxiv.org/abs/1207.1158}{{\tt arXiv:1207.1158 [hep-ex]}}.

\bibitem{Abbiendi:2002vz}
{\bf OPAL} Collaboration, G.~Abbiendi {\em et al.}, ``{Search for nearly mass
  degenerate charginos and neutralinos at LEP},''
  \href{http://dx.doi.org/10.1140/epjc/s2003-01237-x}{{\em Eur. Phys. J.} {\bf
  C29} (2003)  479--489},
\href{http://arxiv.org/abs/hep-ex/0210043}{{\tt arXiv:hep-ex/0210043
  [hep-ex]}}.

\bibitem{Abdallah:2003xe}
{\bf DELPHI} Collaboration, J.~Abdallah {\em et al.}, ``{Searches for
  supersymmetric particles in $e^+ e^-$ collisions up to 208 GeV and
  interpretation of the results within the MSSM},''
  \href{http://dx.doi.org/10.1140/epjc/s2003-01355-5}{{\em Eur. Phys. J.} {\bf
  C31} (2003)  421--479},
\href{http://arxiv.org/abs/hep-ex/0311019}{{\tt arXiv:hep-ex/0311019
  [hep-ex]}}.

\bibitem{Abbiendi:2003sc}
{\bf OPAL} Collaboration, G.~Abbiendi {\em et al.}, ``{Search for chargino and
  neutralino production at $\sqrt{s} = 192$ GeV to 209 GeV at LEP},''
  \href{http://dx.doi.org/10.1140/epjc/s2004-01758-8}{{\em Eur. Phys. J.} {\bf
  C35} (2004)  1--20},
\href{http://arxiv.org/abs/hep-ex/0401026}{{\tt arXiv:hep-ex/0401026
  [hep-ex]}}.

\bibitem{Bramante:2014tba}
J.~Bramante, P.~J. Fox, A.~Martin, B.~Ostdiek, T.~Plehn, T.~Schell, and
  M.~Takeuchi, ``{Relic neutralino surface at a 100 TeV collider},''
  \href{http://dx.doi.org/10.1103/PhysRevD.91.054015}{{\em Phys. Rev.} {\bf
  D91} (2015)  054015},
\href{http://arxiv.org/abs/1412.4789}{{\tt arXiv:1412.4789 [hep-ph]}}.

\bibitem{Boehm:1999bj}
C.~Boehm, A.~Djouadi, and M.~Drees, ``{Light scalar top quarks and
  supersymmetric dark matter},''
  \href{http://dx.doi.org/10.1103/PhysRevD.62.035012}{{\em Phys. Rev.} {\bf
  D62} (2000)  035012},
\href{http://arxiv.org/abs/hep-ph/9911496}{{\tt arXiv:hep-ph/9911496
  [hep-ph]}}.

\bibitem{Kim:2014noa}
J.~S. Kim and T.~S. Ray, ``{The higgsino-singlino world at the large hadron
  collider},'' \href{http://dx.doi.org/10.1140/epjc/s10052-015-3281-4}{{\em
  Eur. Phys. J.} {\bf C75} (2015)  40},
\href{http://arxiv.org/abs/1405.3700}{{\tt arXiv:1405.3700 [hep-ph]}}.

\bibitem{ATLAS-CONF-2015-067}
{\bf ATLAS} Collaboration, ``{Search for pair-production of gluinos decaying
  via stop and sbottom in events with $b$-jets and large missing transverse
  momentum in $\sqrt{s}=13$ TeV $pp$ collisions with the ATLAS detector},''
  Tech. Rep. ATLAS-CONF-2015-067, CERN, Geneva, Dec, 2015.
\newblock \url{http://cds.cern.ch/record/2114839}.

\bibitem{Read:2002hq}
A.~L. Read, ``{Presentation of search results: The CL(s) technique},''
\href{http://dx.doi.org/10.1088/0954-3899/28/10/313}{{\em J. Phys.} {\bf G28}
  (2002)  2693--2704}.

\bibitem{Linnemann:2003vw}
J.~T. Linnemann, ``{Measures of significance in HEP and astrophysics},'' {\em
  eConf} {\bf C030908} (2003)  MOBT001,
\href{http://arxiv.org/abs/physics/0312059}{{\tt arXiv:physics/0312059
  [physics.data-an]}}.

\bibitem{ATL-PHYS-PUB-2014-010}
{\bf ATLAS} Collaboration, ``{Search for Supersymmetry at the high luminosity
  LHC with the ATLAS experiment},'' Tech. Rep. ATL-PHYS-PUB-2014-010, CERN,
  Geneva, Jul, 2014.
\newblock \url{http://cds.cern.ch/record/1735031}.

\bibitem{ATL-PHYS-PUB-2013-011}
{\bf ATLAS} Collaboration, ``{Prospects for benchmark Supersymmetry searches at
  the high luminosity LHC with the ATLAS Detector},'' Tech. Rep.
  ATL-PHYS-PUB-2013-011, CERN, Geneva, Sep, 2013.
\newblock \url{http://cds.cern.ch/record/1604505}.

\bibitem{Alwall:2011uj}
J.~Alwall, M.~Herquet, F.~Maltoni, O.~Mattelaer, and T.~Stelzer, ``{MadGraph 5
  : Going Beyond},'' \href{http://dx.doi.org/10.1007/JHEP06(2011)128}{{\em
  JHEP} {\bf 06} (2011)  128},
\href{http://arxiv.org/abs/1106.0522}{{\tt arXiv:1106.0522 [hep-ph]}}.

\bibitem{Alwall:2014hca}
J.~Alwall, R.~Frederix, S.~Frixione, V.~Hirschi, F.~Maltoni, O.~Mattelaer,
  H.~S. Shao, T.~Stelzer, P.~Torrielli, and M.~Zaro, ``{The automated
  computation of tree-level and next-to-leading order differential cross
  sections, and their matching to parton shower simulations},''
  \href{http://dx.doi.org/10.1007/JHEP07(2014)079}{{\em JHEP} {\bf 07} (2014)
  079},
\href{http://arxiv.org/abs/1405.0301}{{\tt arXiv:1405.0301 [hep-ph]}}.

\bibitem{Aad:2015zva}
{\bf ATLAS} Collaboration, G.~Aad {\em et al.}, ``{Search for new phenomena in
  final states with an energetic jet and large missing transverse momentum in
  pp collisions at $\sqrt{s}=8$ TeV with the ATLAS detector},''
  \href{http://dx.doi.org/10.1140/epjc/s10052-015-3517-3,
  10.1140/epjc/s10052-015-3639-7}{{\em Eur. Phys. J.} {\bf C75} (2015) no.~7,
  299}, \href{http://arxiv.org/abs/1502.01518}{{\tt arXiv:1502.01518
  [hep-ex]}}.
[Erratum: Eur. Phys. J.C75,no.9,408(2015)].

\bibitem{Aad:2013ija}
{\bf ATLAS} Collaboration, G.~Aad {\em et al.}, ``{Search for direct
  third-generation squark pair production in final states with missing
  transverse momentum and two $b$-jets in $\sqrt{s} =$ 8 TeV $pp$ collisions
  with the ATLAS detector},''
  \href{http://dx.doi.org/10.1007/JHEP10(2013)189}{{\em JHEP} {\bf 1310} (2013)
   189},
\href{http://arxiv.org/abs/1308.2631}{{\tt arXiv:1308.2631 [hep-ex]}}.

\bibitem{Aad:2014wea}
{\bf ATLAS} Collaboration, G.~Aad {\em et al.}, ``{Search for squarks and
  gluinos with the ATLAS detector in final states with jets and missing
  transverse momentum using $\sqrt{s}=8$ TeV proton--proton collision data},''
  \href{http://dx.doi.org/10.1007/JHEP09(2014)176}{{\em JHEP} {\bf 09} (2014)
  176},
\href{http://arxiv.org/abs/1405.7875}{{\tt arXiv:1405.7875 [hep-ex]}}.

\bibitem{Aad:2014lra}
{\bf ATLAS} Collaboration, G.~Aad {\em et al.}, ``{Search for strong production
  of supersymmetric particles in final states with missing transverse momentum
  and at least three $b$-jets at $\sqrt{s}=$ 8 TeV proton-proton collisions
  with the ATLAS detector},''
  \href{http://dx.doi.org/10.1007/JHEP10(2014)024}{{\em JHEP} {\bf 10} (2014)
  024},
\href{http://arxiv.org/abs/1407.0600}{{\tt arXiv:1407.0600 [hep-ex]}}.

\bibitem{LeCompte:2011cn}
T.~J. LeCompte and S.~P. Martin, ``{Large Hadron Collider reach for
  supersymmetric models with compressed mass spectra},''
  \href{http://dx.doi.org/10.1103/PhysRevD.84.015004}{{\em Phys. Rev.} {\bf
  D84} (2011)  015004},
\href{http://arxiv.org/abs/1105.4304}{{\tt arXiv:1105.4304 [hep-ph]}}.

\bibitem{Dreiner:2012gx}
H.~K. Dreiner, M.~Kramer, and J.~Tattersall, ``{How low can SUSY go? Matching,
  monojets and compressed spectra},''
  \href{http://dx.doi.org/10.1209/0295-5075/99/61001}{{\em Europhys. Lett.}
  {\bf 99} (2012)  61001},
\href{http://arxiv.org/abs/1207.1613}{{\tt arXiv:1207.1613 [hep-ph]}}.

\bibitem{Dreiner:2012sh}
H.~Dreiner, M.~Kramer, and J.~Tattersall, ``{Exploring QCD uncertainties when
  setting limits on compressed supersymmetric spectra},''
  \href{http://dx.doi.org/10.1103/PhysRevD.87.035006}{{\em Phys.Rev.} {\bf D87}
  (2013) no.~3, 035006},
\href{http://arxiv.org/abs/1211.4981}{{\tt arXiv:1211.4981 [hep-ph]}}.

\bibitem{Carena:2008mj}
M.~Carena, A.~Freitas, and C.~E.~M. Wagner, ``{Light Stop Searches at the LHC
  in Events with One Hard Photon or Jet and Missing Energy},''
  \href{http://dx.doi.org/10.1088/1126-6708/2008/10/109}{{\em JHEP} {\bf 10}
  (2008)  109},
\href{http://arxiv.org/abs/0808.2298}{{\tt arXiv:0808.2298 [hep-ph]}}.

\bibitem{Drees:2012dd}
M.~Drees, M.~Hanussek, and J.~S. Kim, ``{Light Stop Searches at the LHC with
  Monojet Events},'' \href{http://dx.doi.org/10.1103/PhysRevD.86.035024}{{\em
  Phys. Rev.} {\bf D86} (2012)  035024},
\href{http://arxiv.org/abs/1201.5714}{{\tt arXiv:1201.5714 [hep-ph]}}.

\bibitem{Bornhauser:2010mw}
S.~Bornhauser, M.~Drees, S.~Grab, and J.~S. Kim, ``{Light Stop Searches at the
  LHC in Events with two b-Jets and Missing Energy},''
  \href{http://dx.doi.org/10.1103/PhysRevD.83.035008}{{\em Phys. Rev.} {\bf
  D83} (2011)  035008},
\href{http://arxiv.org/abs/1011.5508}{{\tt arXiv:1011.5508 [hep-ph]}}.

\bibitem{Hiller:2009ii}
G.~Hiller, J.~S. Kim, and H.~Sedello, ``{Collider Signatures of Minimal Flavor
  Mixing from Stop Decay Length Measurements},''
  \href{http://dx.doi.org/10.1103/PhysRevD.80.115016}{{\em Phys. Rev.} {\bf
  D80} (2009)  115016},
\href{http://arxiv.org/abs/0910.2124}{{\tt arXiv:0910.2124 [hep-ph]}}.

\bibitem{Aad:2013gva}
{\bf ATLAS} Collaboration, G.~Aad {\em et al.}, ``{Search for long-lived
  stopped R-hadrons decaying out-of-time with pp collisions using the ATLAS
  detector},'' \href{http://dx.doi.org/10.1103/PhysRevD.88.112003}{{\em Phys.
  Rev.} {\bf D88} (2013) no.~11, 112003},
\href{http://arxiv.org/abs/1310.6584}{{\tt arXiv:1310.6584 [hep-ex]}}.

\bibitem{Chalons:2015vja}
G.~Chalons and D.~Sengupta, ``{Closing in on compressed gluino-neutralino
  spectra at the LHC},'' \href{http://dx.doi.org/10.1007/JHEP12(2015)129}{{\em
  JHEP} {\bf 12} (2015)  129},
\href{http://arxiv.org/abs/1508.06735}{{\tt arXiv:1508.06735 [hep-ph]}}.

\end{thebibliography}\endgroup

 
\end{document}